\documentclass[aps,prd,groupedaddress,showpacs,showkeys]{revtex4}
\usepackage{amssymb}
\usepackage{amsmath}
\usepackage{amsfonts}
\usepackage{epsfig}
\usepackage{graphicx}
\usepackage{tabularx}

\newcommand{\beaa}{\begin{eqnarray*}}
\newcommand{\enaa}{\end{eqnarray*}}
\newcommand{\bea}{\begin{eqnarray}}
\newcommand{\ena}{\end{eqnarray}}
\newcommand{\seq}{\begin{subequations}}
\newcommand{\sen}{\end{subequations}}
\newcommand{\eq}{\begin{eqnarray}}
\newcommand{\en}{\end{eqnarray}}

\def\shiftdown#1{#1\llap{\lower.04ex\hbox{#1}}}

\newcommand{\ra}{\rangle}
\newcommand{\la}{\langle}

\def\arraystretch{1.5}

\begin{document}

\title{Nucleon structure including high Fock states 
in AdS/QCD} 

\noindent
\author{
Thomas Gutsche$^1$,
Valery E. Lyubovitskij$^1$
\footnote{On leave of absence
from Department of Physics, Tomsk State University,
634050 Tomsk, Russia},
Ivan Schmidt$^2$,
Alfredo Vega$^2$
\vspace*{1.2\baselineskip}\\
}

\affiliation{
$^1$ Institut f\"ur Theoretische Physik,
Universit\"at T\"ubingen, \\
Kepler Center for Astro and Particle Physics,
\\ Auf der Morgenstelle 14, D-72076 T\"ubingen, Germany
\vspace*{1.2\baselineskip} \\
\hspace*{-1cm}
$^2$ Departamento de F\'\i sica y Centro Cient\'\i fico 
Tecnol\'ogico de Valpara\'\i so (CCTVal), Universidad T\'ecnica
Federico Santa Mar\'\i a, Casilla 110-V, Valpara\'\i so, Chile
\vspace*{1.2\baselineskip}\\}

\date{\today}

\begin{abstract}

We present a detailed analysis of nucleon electromagnetic
and axial form factors in a holographic soft-wall model.
This approach is based on an action which describes
hadrons with broken conformal invariance and incorporates
confinement through the presence of a background dilaton field.
For $N_c=3$ we describe the nucleon structure in a superposition 
of a three valence quark state with high Fock states including 
an adjustable number of partons (quarks, antiquarks and gluons) 
via studying the dynamics of 5D fermion fields 
of different scaling dimension in anti-de Sitter (ADS) space.  
According to the gauge/gravity duality the 5D fermion fields of different 
scaling dimension correspond to the Fock state components with a specific 
number of partons. In the present application we restrict
to the contribution of 3, 4 and 5 parton components in the nucleon Fock state.
With a minimal number of free parameters (dilaton scale parameter, mixing 
parameters of partial contributions of Fock states, coupling 
constants in the effective Lagrangian) we achieve a reasonable agreement with 
data for the nucleon form factors. 

\end{abstract}



\maketitle

\section{Introduction}

Based on the gauge/gravity duality~\cite{Maldacena:1997re},
a class of AdS/QCD approaches which model QCD by using methods of
extra-dimensional field theories formulated in anti-de Sitter (AdS)
space, was recently successfully developed for describing the
phenomenology of hadronic properties (for a recent review see 
e.g.~\cite{Kim:2011ey}). One of the popular formalisms of this kind 
is the ``soft-wall'' model~\cite{Soft_wall1}-\cite{Gutsche:2011vb}, 
which uses a soft infrared (IR) cutoff in the fifth dimension. 
This procedure can be introduced in the
following ways: i) as a background field (dilaton) in the overall
exponential of the action (``dilaton'' soft-wall model),
ii) in the warping factor of the AdS metric (``metric'' soft-wall model), 
iii) in the effective potential of the action. 
In Ref.~\cite{Gutsche:2011vb} we showed that these three ways of proceeding 
are equivalent to each other via a redefinition of the bulk fields and 
by inclusion of extra effective potentials in the action. In our opinion, 
the "dilaton" form of the soft-wall model is more convenient in performing 
the calculations. 

Applications of the soft-wall model to baryon physics have been worked
out in Refs.~\cite{Soft_wall3b,Soft_wall6,Soft_wall5b,Soft_wall9,Soft_wall7,%
Gutsche:2011vb,Vega:2012iz}, where the 
mass spectrum of light and heavy baryons, and 
electromagnetic and gravitational form factors have been calculated.   
We should stress that during last years significant progress in 
understanding of baryon structure using methods of AdS/QCD 
has been achieved~\cite{Soft_wall7,deTeramond:2005su,%
Baryons_ADS_QCD1,Baryons_ADS_QCD2,HW_ADS_QCD2,ADS_QCD_String1,ADS_QCD_String2,%
Henningson:1998cd}.  
In particular, different types of hard-wall models have been suggested 
and developed in Refs.~\cite{Soft_wall7,deTeramond:2005su,%
Baryons_ADS_QCD1,Baryons_ADS_QCD2}.  
Solitonic approaches, where 
stable solitons arise from an effective mesonic action 
which are 5D analogues of 4D Skyrmions, have been suggested in 
Refs.~\cite{HW_ADS_QCD2}. Direct derivations of holographic solitonic 
approaches for baryons from string theories have been proposed in 
Refs.~\cite{ADS_QCD_String1,ADS_QCD_String2}.  
In reference to the 5D soliton AdS/QCD models developed in 
Refs.~\cite{HW_ADS_QCD2,ADS_QCD_String1,ADS_QCD_String2}
we view our approach as an effective or phenomenological framework 
describing baryons in terms of fermion fields. 
As stressed in Ref.~\cite{Baryons_ADS_QCD1} this is not in contradiction 
with basic principles of QCD, because in 4D QCD the baryons can be described 
as skyrmions of the chiral meson Lagrangian or equivalently in terms of 
separate fermion fields coupled to mesons. 
In our approach the fermion bulk fields are characterized by 
the 5D mass $\mu$ (scaling dimension), which is holographically dual to 
$N$ -- the number of partons in baryons. Both quantities scale in 
the large $N_c$ expansion as $\mu \sim N \sim N_c$, which means that the 
baryon is a bound state of $N_c$ quarks. This is consistent with 
large $N_c$ QCD. On the other hand, keeping in mind that in QCD the number 
of colors is equal $N_c = 3$, we in physical applications identify the AdS 
fermion field of lowest dimension with the baryons containing three quarks. 
We do not restrict to the three-valence quark picture of baryons and also 
include higher Fock states involving nonvalence degrees of freedom. 
The latter are dual to the AdS fermion fields of higher dimension.

Here we present a detailed analysis of the nucleon electromagnetic
form factors in a holographic soft-wall model considering the inclusion of 
higher-dimensional fermion fields. Thus high-Fock state contributions
are holographically incorporated in the nucleon.
This novel approach is based on an action which describes
hadrons with broken conformal invariance and which incorporates
confinement through the presence of a background dilaton field. 
Notice that the role of higher Fock components in the pion, 
in the context of holographic QCD, was considered before in 
Refs.~\cite{deTeramond:2012xk,Brodsky:2011xx}. In particular, two 
Fock components 
($q\bar q$ and $q\bar q q \bar q$) were included in the 
expansion of the pion wave function, which was in turn used in the calculation 
of pion electromagnetic and $\gamma\gamma^\ast \pi^0$ transition form factors. 
It was argued that the components containing gluons (e.g. 
$q\bar q g$) are absent in the confinement potential. 

In our framework for $N_c = 3$ nucleons are considered as a superposition 
of three valence quark states and high Fock states including an adjustable 
number of partons (quarks, antiquarks and gluons) by studying the dynamics 
of the 5D fermion fields of different scaling dimension 
in anti-de Sitter (ADS) space.  
According to the gauge/gravity duality the 5D fermion fields of different 
scaling dimension correspond to Fock state components with a specific 
number of partons. 
We can sum the bulk fermion actions with an adjustable 5D fermion mass, which
is related to the scaling dimension (or the number of partons in the nucleon). 
This action is consistent with $C$--, $P$-- and $T$--invariance. Also,
electromagnetic gauge invariance is fulfilled. 
Therefore, the main advantage of our approach is that it allows to include
dynamically any adjustable number of higher Fock states in the nucleon.  
For this first time we restrict ourselves to the contribution of 3, 4 and 
5 parton components in the nucleon Fock state.

The paper is structured as follows. First, in
Sec.~II, we briefly discuss the basic notions of the approach. In
Sec.~III, we consider applications of our approach to the electromagnetic 
properties of the nucleon. Finally, in Sec.~IV, we summarize our results.

\section{Approach} 

We consider the propagation of a fermion field $\Psi(x,z)$ with spin $J=1/2$ 
in 5-dimensional AdS space, which contains the contributions of different 
twist-dimensions. In the language of the AdS/QCD dictionary it corresponds to 
the inclusion of the three-quark and higher-parton states in the nucleon. 
For this first time we restrict ourselves to the contribution of 
$3q$, $3q+g$, $3q+q\bar q$ and $3q+2g$ Fock states, where $q$, $\bar q$ 
and $g$ denote quark, antiquark and gluon, respectively.  

The AdS metric is specified by
\eq
ds^2 = 
g_{MN} dx^M dx^N &=& \eta_{ab} \, e^{2A(z)} \, dx^a dx^b = e^{2A(z)}
\, (\eta_{\mu\nu} dx^\mu dx^\nu - dz^2)\,, 
\nonumber\\
\eta_{\mu\nu} &=& {\rm diag}(1, -1, -1, -1, -1) \,,
\en
where $M$ and $N = 0, 1, \cdots , 4$ are the space-time (base manifold) 
indices, $a=(\mu,z)$ and $b=(\nu,z)$ are the local Lorentz (tangent) indices,
and $g_{MN}$ and  $\eta_{ab}$ are curved and flat metric tensors, which
are related by the vielbein $\epsilon_M^a(z)= e^{A(z)} \, \delta_M^a$ as 
$g_{MN} =\epsilon_M^a \epsilon_N^b \eta_{ab}$. Here
$z$ is the holographic coordinate, $R$ is the AdS radius, and $g =
|{\rm det} g_{MN}| = e^{10 A(z)}$. In the following we restrict
ourselves to a conformal-invariant metric with $A(z) = \log(R/z)$.

The main idea for describing the nucleon in AdS/QCD is based on the 
correspondence (see detailed discussion in 
Refs.\cite{Henningson:1998cd,Baryons_ADS_QCD1,Baryons_ADS_QCD2})  
between the spinor fields propagating in the bulk space and the QCD 
interpolating operators creating the nucleons on the boundary of AdS space. 
The appropriate boundary conditions for the bulk field on the boundary 
of AdS space ensure that such correspondence is precise due to the
equivalence of 
the functional integrals of both the boundary and bulk theories. 
In particular, in the boundary theory (QCD) we define the left- and 
right-handed chiral doublets of nucleons ${\cal O}_L = (p_L, n_L)^T$ and 
${\cal O}_R = (p_R, n_R)^T$, which are fundamental representations  
of the chiral $SU_L(2)$ and $SU_R(2)$ subgroups. Since the chiral symmetry 
of the boundary theory is equivalent to the gauge symmetry in the bulk, 
we need to introduce the pair of bulk fermion fields $\Psi_\pm(x,z)$, 
which are holographic analogues of the 
${\cal O}_{R/L}$ operators. 
In particular, the bulk fields $\Psi_\pm(x,z)$ contain 
important information about the baryon structure. 
On one side, their boundary 
values (non-normalizable solutions) are analogues of the sources for the 
QCD interpolating operators, which then via the evaluation of the Euclidean 
generating functionals produce the correlation functions of 
QCD operators. On the other side, these fields contain normalizable 
modes (these are regular and therefore are vanishing on the boundary) 
- profiles in extra dimension, 
which correspond to the baryon wave functions or expectation values of QCD 
operators. 
In our approach the conformal and chiral symmetries are 
spontaneously broken via the introduction of the background field (dilaton) 
$\varphi(z)$ in the effective action. We choose the quadratic dependence 
of the dilaton on the holographic coordinate $z$, i.e. 
$\varphi(z) = \kappa^2 z^2$ with $\kappa$ being a free scale  parameter, 
which scales as ${\cal O}(\sqrt{N_c})$ in the large $N_c$-expansion.   
In particular, later we show that the nucleon (baryon) mass is 
proportional to the parameter $\kappa$, which is consistent with 
large $N_c$ QCD: $M_N \sim \kappa \sqrt{N_c} \sim N_c$. 
The dilaton can be considered as the expectation 
value of the scalar bulk field with dimension $2$, which is holographically 
dual to the dimension-2 gluon operator $A_\mu^2$.    
Therefore, $\kappa^2$  is related to the vacuum expectation value (VEV)   
$\la \alpha_s A_\mu^2 \ra \sim N_c$ and scales as $\kappa \sim \sqrt{N_c}$. 
Note, the dimension-2 gluon operator $A_\mu^2$ 
has been discussed in the literature (see e.g. 
Refs.~\cite{Celenza:1986th}-\cite{RuizArriola:2004en}). 
The interpretation of the 
dilaton as the quantity dual to the condensate of the dimension-2 operator 
has been done in the framework of the soft-wall model~\cite{Soft_wall3a} 
where the dilaton was introduced in the warping 
factor, breaking the conformal-invariant background metric. 
The main advantage of the dilaton with quadratic profile is the possibility 
to produce linear Regge-like trajectories for hadron masses.  
On the other hand, a quadratic form of the dilaton profile 
is not unique. For example, in the Liu-Tseytlin 
model (a type of top-down AdS/QCD approach)~\cite{Liu:1999fc}, 
the conformal invariance is 
violated by the dilaton, taken in the form $e^{\varphi(z)} = 1 + q z^4$.
The parameter $q$, according to the AdS/QCD 
dictionary~\cite{Klebanov:1999tb}, is related to 
the matrix element of a QCD operator: in particular, 
the scalar $\la \alpha_s G_{\mu\nu}^2 \ra \sim N_c$ and pseudoscalar 
$\la \alpha_s G_{\mu\nu} \tilde G_{\mu\nu} \ra \sim N_c$ gluon condensates. 

An additional source for the breaking of chiral and conformal 
symmetries is the coupling of the $\Psi_+(x,z)$ and $\Psi_-(x,z)$ fields, 
which is an essential basic block of hard-wall AdS/QCD approaches. 
In latter models the conformal invariance is broken by the introduction of an
infrared brane, cutting the AdS geometry in the $z$-direction. 
In addition the couplings of the $\Psi_+(x,z)$ and $\Psi_-(x,z)$ 
fields with the scalar bulk field of dimension-3 are included. Due to
the existence of a
VEV of this scalar field, the chiral and conformal invariances 
are broken. In our approach such a mechanism could in principle be included, 
however it is a higher-order effect since it is generated by 
operators of dimension higher than 2. In particular, 
an extra power of the holographic coordinate gives 
the extra power of $1/\kappa$, which scales as $1/\sqrt{N_c}$. 
In Appendix A we explicitly demonstrate how the coupling between 
$\Psi_+(x,z)$ and $\Psi_-(x,z)$ fields modifies our formalism. 
In the following consideration (including the physical applications) 
we neglect such a coupling.  

The relevant AdS/QCD action for the description of the nucleon electromagnetic 
and axial form factors is constructed in terms of the fermion fields 
$\Psi_{\pm, \tau}(x,z)$ with spin $J=1/2$ and scaling dimension $\tau$ 
(the isospin index corresponding to the proton and neutron components 
is suppressed for simplicity),
the vector field $V_M(x,z)$ with spin $J=1$ (holographic analogue of 
the electromagnetic field)
and the axial field $A_M(x,z)$ (holographic analogue of the axial isovector 
field)~\cite{Soft_wall6,Soft_wall7,Soft_wall9,Gutsche:2011vb}:    
\eq\label{eff_action} 
S &=&  \int d^4x dz \, \sqrt{g} \, e^{-\varphi(z)} \, 
\biggl\{ {\cal L}_\Psi(x,z) + {\cal L}_{V+A}(x,z) 
+ {\cal L}_{\rm int}(x,z) \biggr\} \,, \nonumber\\  
{\cal L}_\Psi(x,z) &=& \sum\limits_{i=+,-} 
\sum\limits_\tau \, c_\tau \, \bar\Psi_{i,\tau}(x,z) 
\, \hat{\cal D}_i(z) \, \Psi_{i,\tau}(x,z) \,, \nonumber\\
{\cal L}_{V+A}(x,z)    &=& - \frac{1}{4} V_{MN}(x,z)V^{MN}(x,z) 
- \frac{1}{4} A_{MN}(x,z)A^{MN}(x,z) \,, \nonumber\\
{\cal L}_{\rm int}(x,z) &=& \sum\limits_{i=+,-} 
\sum\limits_{\tau} \, c_\tau \, 
\bar\Psi_{i, \tau}(x,z) \, \biggl\{ \, \hat{\cal V}_i(x,z) \, 
+ \hat{\cal A}_i(x,z) \, \biggr\}
\, \Psi_{i, \tau}(x,z)\,, 
\en 
where 
\eq 
\hat{\cal D}_\pm(z) &=&  \frac{i}{2} \Gamma^M 
\! \stackrel{\leftrightarrow}{\partial}_{_M} 
\, \mp \,  (\mu + U_F(z))\,, \nonumber\\
\hat{\cal V}_\pm(x,z)  &=&  Q \, \Gamma^M  V_M(x,z) \, \pm \, 
\frac{i}{4} \, \eta_V \,  [\Gamma^M, \Gamma^N] \, V_{MN}(x,z)  
\, \pm \, g_V\, \tau_3 \, \Gamma^M \, i\Gamma^z \, V_M(x,z)  
\,,  \nonumber\\
\hat{\cal A}_\pm(x,z)  &=&  \frac{\tau_3}{2} \, 
\Big (\mp \Gamma^M  A_M(x,z) \, + \, 
\frac{i}{4} \, \eta_A \,  [\Gamma^M, \Gamma^N] \, A_{MN}(x,z)  
\, + \, g_A \, \Gamma^M \, i\Gamma^z \, A_M(x,z) \Big)  
\,. 
\en
Here 
$F_{MN} = \partial_M F_N - \partial_N F_M$ $(F=V,A)$ is the stress tensor 
of the vector (axial) field, $Q = {\rm diag}(1, 0)$ is the nucleon charge 
matrix, $\tau_3 = {\rm diag}(1, -1)$ is the Pauli isospin matrix, 
$A \! \stackrel{\leftrightarrow}{\partial} \! B 
\equiv A (\partial B) - (\partial A) B$, 
$\varphi(z) = \kappa^2 z^2$ is the dilaton field with $\kappa$ being 
a free scale parameter. 
$\Gamma^M = \epsilon_a^M \Gamma^a$ and  
$\Gamma^a=(\gamma^\mu, - i\gamma^5)$ are the five-dimensional Dirac matrices 
(we use the chiral representation for the 
$\gamma^\mu$ and $\gamma^5$ matrices; see details in Appendix B).  
Note that the non-abelian part of the action is irrelevant for the results 
predicted in the paper: the mass spectrum, the electromagnetic and axial 
isovector form factors of nucleons. 
The quantity $\mu$ is the bulk fermion mass related to the scaling dimension 
$\tau$ as $m = \mu R = \tau - 3/2$. Notice that the scaling
dimension of the AdS fermion field is holographically identified 
with the scaling dimension of the baryon interpolating
operator $\tau = N + L$, 
where $N$ is the number of partons in the baryon and 
$L = {\rm max} \, | L_z |$ is the maximal 
value of the $z$-component of the quark orbital angular momentum 
in the light-front wavefunction~\cite{Soft_wall2aa,Soft_wall2b}. 
In the following we restrict to the ground state of nucleons with $L=0$. 
$U_F(z) = \varphi(z)/R$ is the dilaton field dependent effective potential. 
Its presence is necessary due to the following reason. 
The form of the potential $U_F(z)$ is constrained in order to
get solutions of the EOMs for the fermionic Kaluza-Klein (KK) modes
of left and right chirality, and to have the correct asymptotics of the nucleon
electromagnetic form factors at large 
$Q^2$~\cite{Soft_wall6,Soft_wall7,Soft_wall9}. 

Notice that the fermion masses $m$ and effective potentials $U_F(z)$ 
corresponding to the fields $\Psi_+$ and $\Psi_-$ have opposite 
signs according to the $P$-parity transformation (see details in Appendix B). 
In particular, the absolute sign of the fermion mass is related to the 
chirality of the boundary operator~\cite{Baryons_ADS_QCD1,Baryons_ADS_QCD2}.  
According to our conventions the QCD operators ${\cal O}_R$ and ${\cal O}_L$ 
have positive and negative chirality, and therefore the mass 
terms of the bulk fields $\Psi_+$ and $\Psi_-$ have absolute 
signs ``plus'' and ``minus'', respectively. 
In Refs.~\cite{Baryons_ADS_QCD1,Baryons_ADS_QCD2} a different convention 
for left- and right-handed Weyl spinors was used, which is of course 
irrelevant for observable properties. 
In addition to the minimal coupling of the fermion with the vector and
the axial fields, we also include other possible (nonminimal) couplings.
In particular we introduce: 1) a nonminimal coupling of fermion and 
vector fields in order to generate the Pauli form factors of the nucleon; 
2) a minimal-type coupling, which is absent in four dimensions, 
but exists in five dimensions. In Appendix B we explicitly demonstrate  
that these couplings are consistent with $P$--, $C$-- and $T$--parity 
conservation.  
We will show that these terms do not renormalize the electric 
charge of the bulk fields and contribute only to the $Q^2$-dependence 
or to the slopes of the Dirac nucleon form factors. 
The coupling $g_V$ is a free parameter 
which is not constrained by gauge invariance or discrete symmetries 
($P$--, $C$-- or $T$-parity conservation).  
We will fix these terms by improving the description of the
electromagnetic nucleon radii. 
The diagonal matrix $\eta_V = {\rm diag}\{\eta_V^p,\eta_V^n\}$
contains the coupling constrained by the  
anomalous magnetic moments $k_{p,n}$ of the nucleons
($k_p = 1.793$ and $k_n = -1.913$ are given in units of 
nucleon magnetons or n.m.) as 
$\eta^{p,n}_V \sim k_{p,n} \cdot \kappa/m_N$ where 
$m_N$ is the nucleon mass. 

In the case of the axial field we additionally include: 
1) the nonminimal coupling of fermion and axial fields, 
which does not renormalize the axial charge, but gives a 
nontrivial contribution to the $Q^2$-dependence and to the slope 
of the corresponding axial isovector form factor of the nucleon; 
2) the axial-type coupling proportional to the nucleon charge, 
which defines the leading contribution to the isovector 
axial form factor of the nucleon.  

The fields $\Psi_\tau$ describe the AdS fermion 
field with different scaling dimension $\tau$, which in the large $N_c$ 
expansion scales as $\tau \sim N_c$. Restricting to a finite number of 
colors $N_c=3$, we use $\tau = 3, 4, 5$, etc. 
In this paper we restrict to the three leading contributions $\tau = 3, 4$ and 
$5$. According to the AdS/QCD dictionary the fermion field $\Psi_{\tau = 3}$ 
is the holographic analogue of the nucleon interpolating operator with 
twist-dimension 3, which means that the corresponding nucleon Fock state 
contains 3 valence quarks. The fermion field $\Psi_{\tau = 4}$ effectively 
models the nucleon operator with twist-4 (the corresponding 
Fock state contains 4 partons -- 3 valence quarks plus a gluon field). 
Finally, the fermion field $\Psi_{\tau = 5}$ models the nucleon operators 
with twist-5. The corresponding Fock states contain 5 partons: 
(1) 3 valence quarks plus a $q \bar q$ pair of sea quarks or 
(2) 3 valence quarks plus 2 gluons. 
Therefore, the coefficients $c_\tau$ are a set of parameters which take
into account the mixing of AdS fermion fields  with different scaling
dimension $\tau$. 
The set of mixing parameters $c_\tau$ is constrained by 
the correct normalization of the kinetic term of the four-dimensional spinor 
field and by charge conservation as 
$\sum_\tau \, c_\tau = 1$ (see details below). 
In the consideration of the vector (axial) field we apply the axial gauge 
$V(A)_z(x,z) = 0$.

In Figs. 1-3 we give an illustration for the inclusion of twist-3 (Fig.1), 
twist-4 (Fig.2) and twist-5 (Fig.3) partonic Fock states in the description 
of electromagnetic transition between nucleons. 
Due do the gauge/gravity duality we identify the respective sets of
QCD diagrams to the corresponding vector-current 
transition matrix elements involving the fermion field of corresponding 
twist-dimension. One should stress that AdS/QCD gives a unique possibility 
to describe a set of QCD diagrams just by one graph (for each partonic 
content of the nucleon) and obtain predictions for hadronic observables 
in analytical form. 

Finalizing our discussion of the 5D effective action~(\ref{eff_action}) 
we would like to point out again that it obeys 
$P$--, $C$-- and $T$--invariance. This action further contains 
new terms describing the interaction of vector and axial fields 
with fermions, which were not considered before
in the context of AdS/QCD. 
These new terms do not renormalize the charge (i.e. vanish at $Q^2=0$),
but they contribute to the $Q^2$--dependence and the slopes of the 
corresponding form factors. Their relevance for giving a sufficient
description of the data will be shown further on.

\subsection{Mass spectrum} 

One advantage of the soft-wall AdS/QCD model is that most of the 
calculations can be done analytically. 
In a first step, 
we show how in this approach the baryon spectrum and 
wave functions are generated. We follow the procedure pursued in
Refs.~\cite{Soft_wall6,Soft_wall7,Soft_wall9,Gutsche:2011vb}.  
Dropping the vector and axial fields, and rescaling the fermionic fields as
\eq
\Psi_{i, \tau}(x,z) = e^{\varphi(z)/2} \psi_{i, \tau}(x,z),
\en
we remove the dilaton field from the overall exponential.
In terms of the field $\psi_\tau(x,z)$ the modified action in 
the Lorentzian signature reads as 
\eq\label{S_F2}
S_0 = \int d^4x dz \, e^{4A(z)} \, \sum\limits_{i=+,-} \sum\limits_\tau 
\, c_\tau \, \bar\psi_{i,\tau}(x,z) \, 
\biggl\{ i\not\!\partial + \gamma^5\partial_z
+ 2 A^\prime(z) \gamma^5
-  \delta_i \frac{e^{A(z)}}{R} \Big(m + \varphi(z)\Big)  \biggr\} 
\, \psi_{i, \tau}(x,z)\,,
\en
where $\not\!\partial = \gamma^\mu \, \partial_\mu$, $\delta_\pm = \pm 1$ 
and the fermion field $\psi_{i, \tau}(x,z)$ satisfies the following equation 
of motion (EOM)~\cite{Soft_wall6,Soft_wall7,Soft_wall9,Gutsche:2011vb}: 
\eq\label{EOM_psi_pm}
\biggl[ i\not\!\partial + \gamma^5\partial_z
+ 2 A^\prime(z) \gamma^5
\mp \frac{e^{A(z)}}{R} \Big(m + \varphi(z)\Big) \biggr] 
\psi_{\pm,\tau}(x,z) = 0\,. 
\en
Based on these solutions the fermionic action should be extended by
an extra term in the ultraviolet boundary (see details in
Refs.~\cite{Soft_wall7,Henningson:1998cd}) in order to guarantee 
the gauge/gravity correspondence --- equivalence between the 
AdS functional integral and the generating functional for correlation 
functions in QCD. 

Next we split the fermion field into left- and right-chirality
components 
\eq
\psi_{i, \tau}(x,z) = \psi^L_{i,\tau}(x,z) + \psi^R_{i, \tau}(x,z)\,, \quad
\psi^{L/R}_{i, \tau}(x,z) = 
\frac{1 \mp \gamma^5}{2} \psi_{i, \tau}(x,z) \,, \quad
\gamma^5 \psi^{L/R}_{i, \tau}(x,z) = \mp \psi^{L/R}_{i, \tau}(x,z) \,.
\en
and perform a KK expansion for the $\psi^{L/R}_{i, \tau}(x,z)$ fields:
\eq
\psi^{L/R}_{i, \tau}(x,z) = \frac{1}{\sqrt{2}} \, \sum\limits_n
\ \psi^{L/R}_n(x) \ F^{L/R}_{i, \tau, n}(z) \,, 
\en
where $\psi^{L/R}_n(x)$ are the four-dimensional boundary 
fields (KK modes). These are Weyl spinors forming the 
Dirac bispinors $\psi_n(x) = \psi^L_n(x) + \psi^R_n(x)$, 
and $F^{L/R}_{i, \tau,n}(z)$ are the normalizable profile  
functions. Due to four-dimensional $P$- and $C$-parity invariance 
the bulk profiles are related as (see details in Appendix B) 
\eq 
F^R_{\pm, \tau, n}(z) = \mp F^L_{\mp, \tau, n}(z)\,. 
\en 
Using this constraint in the following we use the simplified notations: 
\eq 
F^R_{\tau, n}(z) &\equiv& 
F^R_{+, \tau, n}(z) = - F^L_{-, \tau, n}(z)\,, \nonumber\\ 
F^L_{\tau, n}(z) &\equiv& 
F^L_{+, \tau, n}(z) = F^R_{-, \tau, n}(z)\,. 
\en 
Note that the profiles $F^{L/R}_{\tau, n}(z)$ are the 
holographic analogues of the nucleon wave functions with 
specific radial quantum number $n$ and twist dimension $\tau$ 
(the latter corresponds to the specific partonic content of the nucleon Fock 
component), which  satisfy
the two coupled one-dimensional EOMs~\cite{Soft_wall6,Soft_wall7,Soft_wall9}:
\eq
\biggl[\partial_z \pm \frac{e^{A}}{R} \, \Big(m+\varphi\Big)
+ 2 A^\prime \biggr] F^{L/R}_{\tau, n}(z) = \pm M_{n\tau}
F^{R/L}_{\tau, n}(z) \,. 
\en
Therefore, our main idea is to find the solutions for the 
bulk profiles of the AdS field in the $z$--direction, and then calculate 
the physical properties of hadrons. 
After straightforward algebra one can obtain the decoupled EOMs: 
\eq
\biggl[ -\partial_z^2 - 4 A^\prime \partial_z
+ \frac{e^{2A}}{R^2} (m+\varphi)^2
\mp \frac{e^{A}}{R} \Big(A^\prime (m+\varphi) + \varphi^\prime\Big)
- 4 A^{\prime 2} - 2 A^{\prime\prime}
\biggr] F^{L/R}_{\tau, n}(z) = M_{n\tau}^2 F^{L/R}_{\tau, n}(z) \,.
\en
Doing the substitution
\eq
F^{L/R}_{\tau, n}(z) = e^{- 2 A(z)} \, f^{L/R}_{\tau, n}(z)
\en
we derive the Schr\"odinger-type EOM for $f^{L/R}_{\tau, n}(z)$
\eq\label{eq_KK} 
\biggl[ -\partial_z^2
+ \frac{e^{2A}}{R^2} (m+\varphi)^2
\mp \frac{e^{A}}{R} \Big(A^\prime (m+\varphi) +
 \varphi^\prime\Big) \biggr] f^{L/R}_{\tau, n}(z) 
= M_{n\tau}^2 \, f^{L/R}_{\tau, n}(z) \,.
\en
For $A(z)=\log(R/z)$, $\varphi(z)=\kappa^2 z^2$ we get
\eq
\biggl[ -\partial_z^2
+ \kappa^4 z^2 + 2 \kappa^2 \Big(m \mp \frac{1}{2} \Big)
+ \frac{m (m \pm 1)}{z^2} \biggr] f^{L/R}_{\tau, n}(z) = 
M_{n\tau}^2 \, f^{L/R}_{\tau, n}(z),
\en
where
\eq\label{bulk_prof_dilaton}
f^L_{\tau, n}(z) &=& \sqrt{\frac{2\Gamma(n+1)}{\Gamma(n+\tau)}} 
\ \kappa^{\tau}
\ z^{\tau-1/2} \ e^{-\kappa^2 z^2/2} \ L_n^{\tau - 1}(\kappa^2z^2) \,, \\
f^R_{\tau, n}(z) &=& \sqrt{\frac{2\Gamma(n+1)}{\Gamma(n+\tau-1)}} 
\ \kappa^{\tau-1} \ z^{\tau-3/2} \ e^{-\kappa^2 z^2/2} 
\ L_n^{\tau-2}(\kappa^2z^2)
\en
and
\eq
M_{n\tau}^2 = 4 \kappa^2 \Big( n + \tau - 1 \Big)  
\en
with 
\eq\label{norm_cond} 
\int\limits_0^\infty dz \ f^{L/R}_{\tau, n_1}(z) \, 
f^{L/R}_{\tau, n_2}(z) 
\, = \, \delta_{n_1n_2}\; .
\en 
Here 
\eq 
L_n^\tau(x) = \frac{x^{-\tau} e^x}{n!} 
\, \frac{d^n}{dx^n} \Big( e^{-x} x^{\tau+n} \Big)
\en 
are the generalized Laguerre polynomials. 
In above formulas we substituted $m = \tau  -  3/2$. 

One can see that the functions 
$F^{L/R}_{\tau, n}(z) = e^{- 2 A(z)} \, f^{L/R}_{\tau, n}(z)$ 
have the correct scaling behavior for small $z$
\eq
F^L_{\tau, n}(z) \sim z^{\tau+3/2}\,, \quad\quad
F^R_{\tau, n}(z) \sim z^{\tau+1/2}\,
\en 
when identified with the corresponding nucleon wave functions with
twist $\tau$, 
and vanish at large $z$ (confinement). Below, in the discussion of
the electromagnetic properties of the nucleon, we explicitly demonstrate that 
the Dirac and Pauli form factors have the correct scaling dependence at 
large $Q^2$. 

The nucleon mass is identified with the expression 
\eq 
M_n = \sum\limits_\tau \, c_\tau \, M_{n\tau} 
    = 2 \kappa \sum\limits_\tau \, c_\tau \, \sqrt{n + \tau -1} \,. 
\en  
Due to the dilaton the chiral and conformal symmetries are 
spontaneously broken in our approach. 
Switching off the dilaton field (which corresponds to the limit 
$\kappa = 0$) leads to the restoration of the chiral and 
conformal symmetries. In particular, the masses of the bulk profiles 
$f^{L/R}_{\tau, n}(z)$ and the nucleon mass vanish in this limit. 
As we stressed before, the nucleon (baryon) mass is 
proportional to the parameter $\kappa$ and this is consistent with 
large $N_c$ QCD: $M_N \sim \kappa \sqrt{\tau} \sim N_c$, 
where $\tau \sim N_c$. On the other hand, 
it is known that the nucleon (baryon) mass is proportional 
to the quark condensate $|\la\bar q q\ra|$  
(so-called Ioffe formula)~\cite{Ioffe:1982ce}. 
It means that our soft-wall model indicates that there could be  
a relation between condensates of the dimension-2 gluon operator 
${\cal O}_{A^2} = A_\mu^2$  and dimension-3 scalar quark-antiquark operator 
${\cal O}_{\bar q q} = \bar q q$. As we show in Appendix A the 
contribution of the condensate of the dimension-3 scalar bulk field  
into the nucleon mass is suppressed by one order of $N_c$. 
It means that in the dilaton-type of the soft-wall model the dilaton 
gives the leading contribution to the spontaneous breaking of 
chiral symmetry (and therefore to the nucleon mass) in comparison 
with the dimension-3 scalar bulk field. 

Integration over the holographic coordinate $z$, with the use of the 
normalization condition (\ref{norm_cond}) for the profile functions 
$f^{L/R}_{\tau, n}(z)$,  gives a four-dimensional action for the 
fermion field $\psi_n(x) = \psi^L_n(x) + \psi^R_n(x)$:
\eq
S_{0} = \sum\limits_n \, \int d^4x \, 
\bar\psi_n(x) \biggl[ i \not\!\partial - M_{n} \biggr] \psi_n(x) \,.
\en 
This last equation is a manifestation of the gauge-gravity duality.
It explicitly demonstrates
that effective actions for conventional hadrons in 4 dimensions
can be generated from actions for bulk
fields propagating in 5 dimensional AdS space. The effect of the
extra-dimension is encoded in the hadronic mass squared (in our 
case in the nucleon mass $M_n$, where $n$ is the radial quantum 
number), which is the superposition of the solutions of the Schr\"odinger 
equation~(\ref{eq_KK}) for the KK profiles in the extra dimension. 
Notice that the constraint 
$\sum_\tau \, c_\tau = 1$ for the mixing 
parameters $c_\tau$ was essential in order to get 
the correct normalization of the kinetic term 
$\bar\psi_n(x) i\!\!\not\!\partial\psi_n(x)$ 
of the four-dimensional spinor field. 

\subsection{Electromagnetic structure of nucleons} 

The nucleon electromagnetic form factors $F_1^N$ and $F_2^N$ 
($N=p, n$ correspond to proton and neutron) are conventionally 
defined by the matrix element of the electromagnetic current as 
\eq 
\label{EM_current}
\langle p' | J^{\mu}(0) | p \rangle \, = \, 
\bar{u}(p') \, \left[ \gamma^{\mu} F_{1}^N(t) 
+ \frac{i}{2 m_N} \, \sigma^{\mu \nu} \, q_\nu 
F_{2}^N(t) \right] u(p),
\en 
where $q = p' - p$ is the momentum transfer and $t=q^2$; 
$m_N$ is the nucleon mass; and $F_1^N$ and $F_2^N$ are 
the Dirac and Pauli form factors, which 
are normalized to the electric charge $e_N$ and anomalous 
magnetic moment $k_N$ of the corresponding nucleon: 
$F_1^N(0)=e_N$ and $F_2^N(0)=k_N$.  

In our approach the nucleon form factors are generated by the 
action 
\eq 
S_{\rm int}^V =  \int d^4x dz \, \sqrt{g} \, e^{-\varphi(z)} \, 
{\cal L}_{\rm int}^V(x,z)
\en 
containing the minimal and nonminimal couplings of fermion and vector 
AdS fields. The expressions for the Dirac and Pauli nucleon form factors 
are given by: 
\eq 
F_1^p(Q^2) &=& C_1(Q^2) + g_V C_2(Q^2)
+ \eta_V^p C_3(Q^2)\,,\nonumber\\
F_2^p(Q^2) &=& \eta_V^p C_4(Q^2)\,,\nonumber\\
& &\\
F_1^n(Q^2) &=& - g_V C_2(Q^2) + \eta_V^n C_3(Q^2)\,,
\nonumber\\
F_2^n(Q^2) &=& \eta_V^n C_4(Q^2), \nonumber 
\en 
where 
$Q^{2} = - t$ and 
$C_i(Q^2)$ are the structure integrals: 
\eq 
C_{1}(Q^2) &=& \frac{1}{2} \, \int\limits_0^\infty dz \, V(Q,z) 
\ \sum\limits_\tau \, c_\tau \, 
\biggl( [f^L_{\tau}(z)]^2 + [f^R_{\tau}(z)]^2 \biggr)\,, 
\nonumber\\ 
C_{2}(Q^2) &=& \frac{1}{2} \, \int\limits_0^\infty dz \, 
V(Q,z) \ \sum\limits_\tau \, c_\tau \, 
\biggl( [f^R_{\tau}(z)]^2 - [f^L_{\tau}(z)]^2 \biggr)\,, 
\nonumber\\  
C_{3}(Q^2) &=& \frac{1}{2} \, \int\limits_0^\infty dz z \, \partial_z V(Q,z) 
\ \sum\limits_\tau \, c_\tau \, 
\biggl( [f^L_{\tau}(z)]^2 - [f^R_{\tau}(z)]^2 \biggr)\,, 
\nonumber\\  
C_{4}(Q^2) &=& 2 m_N \, \int\limits_0^\infty dz z \, V(Q,z) 
\ \sum\limits_\tau \, c_\tau \, 
f^L_{\tau}(z) f^R_{\tau}(z)\,. \label{Ci} 
\en 
The functions $f^{R/L}_{\tau}(z) \equiv f^{R/L}_{\tau, n=0}(z)$
are the bulk profiles of fermions 
with $n=0$ (corresponding to the ground-state nucleon 
with radial quantum number $n=0$)  
found in the previous subsection:  
\eq
f^L_{\tau}(z) &=& \sqrt{\frac{2}{\Gamma(\tau)}} 
\ \kappa^{\tau}
\ z^{\tau-1/2} \ e^{-\kappa^2 z^2/2} \,, \\
f^R_{\tau}(z) &=& \sqrt{\frac{2}{\Gamma(\tau-1)}} 
\ \kappa^{\tau-1} \ z^{\tau-3/2} \ e^{-\kappa^2 z^2/2} \,. 
\en
$V(Q,z)$ is the bulk-to-boundary propagator of 
the transverse massless vector bulk field (the holographic analogue of 
the electromagnetic field) defined as 
\eq
V_\mu(x,z) = \int \frac{d^4 q}{(2\pi)^4} e^{-iqx} V_\mu(q) V(q,z) 
\en
and obeys the following EOM 
\eq
  \partial_z \biggl( \frac{e^{-\varphi(z)}}{z} \partial_z V(q,z) \biggr)
+ q^2 \, \frac{e^{-\varphi(z)}}{z} V(q,z) = 0 \,,
\en 
which is derived from the action 
\eq 
S_V =  \int d^4x dz \, \sqrt{g} \, e^{-\varphi(z)} \, 
{\cal L}_V(x,z) \,. 
\en 
In the soft-wall model the solution for $V(Q,z)$ is given in analytical 
form in terms of the Gamma $\Gamma(n)$ and Tricomi 
$U(a,b,z)$ functions: 
\eq\label{V}
V(Q,z) = \Gamma\biggl(1 + \frac{Q^{2}}{4 \kappa^{2}}\biggr) 
U\biggl(\frac{Q^{2}}{4 \kappa^{2}}, 0, \kappa^2 z^2\biggr)\;.
\en 
The bulk-to-boundary propagator $V(Q,z)$ obeys the normalization
condition $V(0,z) = 1$ consistent with 
gauge invariance and fulfills the following ultraviolet (UV) and infrared (IR) 
boundary conditions : 
\eq 
V(Q,0) = 1\,, \quad\quad V(Q,\infty) = 0 \,. 
\en 
The UV boundary condition corresponds to the local (structureless) 
coupling of the electromagnetic field to matter fields, while 
the IR boundary condition implies that the vector field
vanishes at $z=\infty$. 

In order to obtain analytical expressions for the functions $C_i(Q^2)$ 
(see Appendix C) 
it is convenient to use the integral representation for $V(Q,z)$ 
introduced in Ref.~\cite{Soft_wall4a} 
\eq 
\label{VInt}
V (Q,z) = \kappa^{2} z^{2} \int_{0}^{1} \frac{dx}{(1-x)^{2}} 
\, x^{\frac{Q^{2}}{4 \kappa^{2}}} \, 
e^{- \displaystyle{\frac{\kappa^2 z^2 x}{1-x} }}\,,   
\en 
where the variable $x$ is equivalent to the light-cone momentum 
fraction~\cite{Soft_wall2ab}. 

There are a few very important properties of the $C_i(Q^2)$ functions. 
At $Q^2 = 0$ they are normalized as  
\eq 
C_1(0) = 1\,, \quad C_2(0) = C_3(0) = 0\,, 
\quad C_4(0) = \frac{2m_N}{\kappa} \sum\limits_\tau \, 
c_\tau \sqrt{\tau-1} \,. 
\en 
The normalizations of $C_i$ $(i=1,2,3)$ are consistent with 
gauge invariance (the charge normalization for nucleons). 
In particular, this means that the proton and neutron Dirac 
form factors are normalized accordingly: 
\eq 
F_1^p(0) = 1, \quad\quad F_1^n(0) = 0 \,. 
\en 
Here we take into account the constraint $\sum_\tau c_\tau = 1$ 
of the mixing parameters $c_\tau$, which is also essential to 
get the correct normalization of the kinetic term of
the four-dimensional spinor 
field on the boundary of AdS space. The anomalous magnetic moments 
of the nucleons $N=p,n$ are given by 
\eq 
\kappa_N = \eta_V^N C_4(0) = 
\frac{2 \eta_V^N m_N}{\kappa} \sum\limits_\tau \, c_\tau \sqrt{\tau-1} \,.
\en 
In the analysis of the electromagnetic form factors we will use 
the dipole formula 
\eq 
G_D(Q^2) = \frac{1}{(1 + Q^2/\Lambda^2)^2} \,,
\en 
where $\Lambda^2 =  0.71$ GeV$^2$. 

\subsection{Axial isovector form factor of nucleons} 

The nucleon isovector axial form factor $G_A(t)$ 
is conventionally defined by the matrix element of the axial 
isovector current as 
\eq 
\label{A_current}
\langle p' | A^{\mu}_3(0) | p \rangle \, = \, 
\bar{u}(p') \, \left[ \gamma^{\mu} G_A(t) 
+ \frac{q^\mu}{2m_N} G_P(t) \right] \, \gamma^5 \, \frac{\tau_3}{2} \ u(p) \,,
\en 
which is normalized to the nucleon axial charge 
$G_A(0)=g_A$. 

In our approach the $G_A(Q^2)$ form factor is generated by the 
action 
\eq 
S_{\rm int}^A =  \int d^4x dz \, \sqrt{g} \, e^{-\varphi(z)} \, 
{\cal L}_{\rm int}^A(x,z)
\en 
containing the minimal and nonminimal couplings of fermion and axial-vector 
AdS fields. The expression for the axial isovector nucleon form factor  
is given by: 
\eq 
G_A(Q^2) &=& g_A D_1(Q^2) + D_2(Q^2) + \eta_A D_3(Q^2) \,,
\en 
where 
$D_i(Q^2)$ are the structure integrals: 
\eq 
D_{1}(Q^2) &=& \frac{1}{2} \, \int\limits_0^\infty dz \, A(Q,z) 
\ \sum\limits_\tau \, c_\tau \, 
\biggl( [f^L_{\tau}(z)]^2 + [f^R_{\tau}(z)]^2 \biggr)\,, 
\nonumber\\ 
D_{2}(Q^2) &=& \frac{1}{2} \, \int\limits_0^\infty dz \, 
A(Q,z) \ \sum\limits_\tau \, c_\tau \, 
\biggl( [f^L_{\tau}(z)]^2 - [f^R_{\tau}(z)]^2 \biggr)\,, 
\nonumber\\  
D_{3}(Q^2) &=& - \frac{1}{2} \, \int\limits_0^\infty dz z \, 
\partial_z A(Q,z) \ \sum\limits_\tau \, c_\tau \, 
\biggl( [f^L_{\tau}(z)]^2 + [f^R_{\tau}(z)]^2 \biggr)\,. 
\label{Di} 
\en 
Now $A(Q,z)$ is the bulk-to-boundary propagator of 
the transverse massless axial bulk field (the holographic analogue of 
the axial isovector field). In our approximation it coincides with 
$V(Q,z)$.   
The functions $D_i(Q^2)$ at $Q^2 = 0$ are normalized as  
\eq 
D_1(0) = 1\,, \quad D_2(0) = D_3(0) = 0\,. 
\en 
As in the case of the $C_i$ functions these results are based on
the normalization 
properties of the bulk profiles $f^{R/L}(z)$ and the constraint condition 
$\sum_\tau c_\tau = 1$. 

Our prediction for the form factor $G_A(Q^2)$ will be compared to 
the dipole fit formula 
\eq 
G_A^D(Q^2) = \frac{g_A}{(1 + Q^2/M_A^2)^2} 
\en 
extracted from neutrino scattering experiments,  
where $M_A = 1.026 \pm 0.021$ GeV~\cite{PDG}.  

\section{Results} 

In this section we present the related numerical analysis of 
nucleon properties: magnetic moments  
($\mu_p = 1 + \kappa_p$, $\mu_n = \kappa_n$), 
electromagnetic and axial radii 
($r_E^p$, $\la r^2_E \ra^n$, $r_M^p$, $r_M^n$, $r_A$), isovector axial 
and electromagnetic Dirac, Pauli and Sachs form factors and their ratios 
in the Euclidean region. 

We first want to recall the definitions of the Sachs form factors 
$G_{E/M}(Q^2)$,  
the electromagnetic $\la r^2_{E/M} \ra^N$ and isovector axial 
$\la r^2_A \ra$ radii: 
\eq 
G_E^N(Q^2) & = & F_1^N(Q^2) - \frac{Q^2}{4m_N^2} F_2^N(Q^2)\,, \quad 
G_M^N(Q^2) \ = \ F_1^N(Q^2) + F_2^N(Q^2)\,, \nonumber\\
\la r^2_E \ra^N & = & - 6 \frac{dG_N^E(Q^2)}{dQ^2}\bigg|_{Q^2 = 0} \,, 
\hspace*{1.6cm} 
\la r^2_M \ra^N \ = \ - \frac{6}{G_M^N(0)} \, 
\frac{dG_M^N(Q^2)}{dQ^2}\bigg|_{Q^2 = 0}  \,, \nonumber\\
\la r^2_A \ra & = & - \frac{6}{G_A(0)} \, 
\frac{dG_A(Q^2)}{dQ^2}\bigg|_{Q^2 = 0} \,, 
\en 
with $G_M^N(0) \equiv \mu_N$ and $G_A(0) \equiv g_A$. 

The five free parameters $\kappa$, $c_3$, $c_4$, $g_V$ and $\eta_A$ 
are fixed to the values 
\eq 
\kappa = 383 \ \mathrm{MeV}\,, \quad c_3 = 1.25\,, \quad c_4 = 0.16\,, 
\quad
g_V = 0.3\,, \quad \eta_A = 0.5\;. 
\en 
Note that the parameter $c_5$ is expressed through $c_3$ and $c_4$ as 
\eq 
c_5 = 1 - c_3 - c_4 = - 0.41 \,. 
\en 
Here the parameters  $c_3$, $c_4$ are constrained by the nucleon mass. 
The parameter $\kappa$ is fixed by the nucleon mass and the
electromagnetic radii. The parameters $g_V$ and $\eta_A$ are 
fitted by fine tuning of the neutron electromagnetic and 
nucleon axial radius, respectively.   
Notice also that the other parameters are fixed by the magnetic moments 
and the axial charge of nucleons and should not be counted as free 
parameters:  
\eq 
g_A = 1.270\,, \quad 
\eta_V^p = \frac{\kappa \, (\mu_p - 1)}{2m_N \, C_0} = 0.30\,, \quad 
\eta_V^n = \frac{\kappa \, \mu_n}{2m_N \, C_0} = - 0.32\,, \quad 
C_0 = \sqrt{2} \,  c_3 + \sqrt{3} \, c_4 + 2 c_5 \;.
\en 
In Table I we present the results for the nucleon mass and the electroweak 
properties of nucleons. 
Results for the nucleon electromagnetic form factors in comparison to
known data are shown in Figs. 4-14. In particular, in Figs.4, 6 
and 7 we present the ratios of proton charge and nucleon magnetic 
form factors to the dipole form factor $G_D$. 
In Fig.5 we present the results for ratio 
$\mu_p \, G_E^p(Q^2)/G_M^p(Q^2)$. In Fig. 8 and 9 we present 
the prediction for the charge neutron form factor and 
the ratio $G_E^n(Q^2)/G_M^n(Q^2)$. In Figs.10 and 11 we present 
the predictions for the Dirac nucleon form factors multiplied by 
$Q^4$. Figs.12 and 13 show the ratios of the Pauli and Dirac form 
factors of the proton multiplied with $Q^2$ and with 
$Q^2/\log^2(Q^2/\Lambda^2)$, where $\Lambda = 0.3$ GeV. 
Finally, in Fig.14 we present the predictions for the ratio 
of nucleon axial isovector form factor to the dipole form factor $G_A^D(Q^2)$. 

We demonstrated that the soft-wall holographic model in the semiclassical
approximation reproduces the main features of the electromagnetic structure
of the nucleon. In particular, we achieved the following results:
the analytical power scaling of the elastic nucleon
form factors at large momentum transfers in accordance with quark-counting
rules; reproduction of experimental data for magnetic moments and
electromagnetic
radii.
 
One can see that with a minimal number of free parameters (five 
parameters) we obtain a reasonable description of the nucleon 
electromagnetic and axial-vector form factors including 
the correct power scaling at large $Q^2$. It demonstrates 
that the soft-wall model successfully describes nucleon structure 
at any resolution scale. 
In a next step, one can include effects of quark masses and 
extend the approach to nucleon resonances, light baryons with 
higher spins, strange and heavy baryons. 

\section{Summary}

We presented a soft-wall model which allows to include higher 
Fock states in the analysis of the nucleon structure. 
This approach is based on an action which describes
hadrons with broken conformal invariance and incorporates
confinement through the presence of a background dilaton field.
For $N_c=3$ 
the nucleon is described in terms of a superposition of the three valence 
quark state with high Fock states with an adjustable number of partons
(quarks, antiquarks and gluons). Its structure is determined by studying the
dynamics of 5D fermion fields of different scaling dimension in 
anti-de Sitter (ADS) space.  
According to the gauge/gravity duality the 5D fermion fields of different 
scaling dimension correspond to Fock state components with a specific 
number of partons. For the first application we restrict ourselves
to the contribution of 3, 4 and 5 parton components in the nucleon Fock state. 
The role of higher Fock components 
in the context of holographic QCD has been already considered in the case 
of the pion~\cite{deTeramond:2012xk,Brodsky:2011xx}. In particular, 
two components ($q\bar q$ and $q\bar q q \bar q$) were included in the 
expansion of the pion wave function, which was then used in the calculation 
of pion electromagnetic and $\gamma\gamma^\ast \pi^0$ transition form factors. 
It was further argued that the components containing gluons (e.g. 
$q\bar q g$) are absent in the confinement potential. In our case the 
contribution of the twist-4 component containing three constituent quarks 
and a single gluon is not zero, but is suppressed, which is partially in line
with the conclusion of Refs.~\cite{deTeramond:2012xk,Brodsky:2011xx}. 
On the other hand, our mechanism generating the inclusion of 
higher Fock states is different from the one suggested 
in~\cite{deTeramond:2012xk,Brodsky:2011xx}. Additionally, 
the pion and the nucleon are quite different hadronic bound states, and 
therefore, the role of Fock states containing gluons 
could be different. We think that this issue requires further investigation. 

We presented a detailed analysis of nucleon electromagnetic
and axial form factors. 
With a minimal number of free parameters (dilaton scale parameter, mixing 
parameters of the partial contributions of Fock states, a few coupling 
constants in the effective Lagrangian) we achieved a reasonable agreement 
with data for the nucleon electromagnetic and axial isovector form factors. 
Note that all form factors have the correct scaling at large $Q^2$. 
As next applications we plan to extend 
our approach to nucleon resonances (e.g. Roper) and baryons with strangeness. 
There is also a possibility to study nuclear systems using methods of 
AdS/QCD via studying dynamics of 5D fields of higher dimensions, which 
holographically correspond to nuclei with a specific number of nucleons 
and electrons.

\begin{acknowledgments}

The authors thank Stan Brodsky, Guy de T\'eramond and Werner Vogelsang 
for useful discussions. 
This work was supported by the DFG under Contract No. LY 114/2-1, 
by Federal Targeted Program "Scientific
and scientific-pedagogical personnel of innovative Russia"
Contract No. 02.740.11.0238, by FONDECYT (Chile) under Grant No. 1100287.
V. E. L. would like to thank Departamento de F\'\i sica y Centro
Cient\'\i fico Tecnol\'ogico de Valpara\'\i so (CCTVal), Universidad
T\'ecnica Federico Santa Mar\'\i a, Valpara\'\i so, Chile for warm
hospitality. A. V. acknowledges the financial support from FONDECYT (Chile)
Grant No. 3100028.

\end{acknowledgments}

\appendix

\section{Coupling of $\Psi_+(x,z)$ and $\Psi_-(x,z)$ fermion fields} 

Following Ref.~\cite{Baryons_ADS_QCD1} we introduce the Yukawa-type 
coupling of $\Psi_+(x,z)$ and $\Psi_-(x,z)$ fields with  
the bulk scalar field $X(x,z)$ dual to the dimension-3 quark operator 
${\cal O}_q = \bar q q$: 
\eq 
{\cal L}_{Y}(x,z) = - \frac{g}{R} \left( \bar\Psi_-(x,z) X(x,z) \Psi_+(x,z)
\, + \, \bar\Psi_+(x,z) X^\dagger(x,z) \Psi_-(x,z) \right)  \,, 
\en 
where $g$ is the coupling constant, which scales in large $N_c$ expansion 
as $g \sim \sqrt{N_c}$. 
As originally was shown, the VEV of the scalar bulk field $X_0(z)$ is 
the linear combination of two solutions (see e.g. Ref.~\cite{Soft_wall1}),  
which for asymptotically small $z$ behaves as: 
\eq\label{S_cond} 
X_0(z) \to \frac{1}{2} \hat{m} z + \frac{1}{2} \Sigma z^3\,,  
\en 
where $\hat{m}$ is the current quark mass and $\Sigma = |\la \bar q q \ra |$ 
is quark condensate in the chiral limit $\hat{m} \to 0$. On the other hand,  
in the original soft-wall model the infrared asymptotics $z \to \infty$ 
dictates that $\Sigma$ is simply proportional to $\hat{m}$ which 
is in contradiction with QCD. It was suggested to include in the effective 
action the potential containing higher-order 
terms in the scalar field to resolve the problem $\Sigma \sim \hat{m}$. 
Later in Ref.~\cite{Gherghetta:2009ac} this idea was realized 
by adding a quartic term $\sim X^4(x,z)$ in the effective action. 
At the same time in Ref.~\cite{Cherman:2008eh} it was noticed that 
the scalar operator ${\cal O}_q$ and its source $J_q$ can always be 
rescaled by a constant $a$  
\eq 
{\cal O}_q \to a \, {\cal O}_q\,, \quad 
J_q \to J_q/a 
\en 
keeping the product $J_q \, {\cal O}_q$ unchanged. Then using 
the arguments of large $N_c$ QCD it was shown that the constant 
$a$ must scale as $a \sim 1/\sqrt{N_c}$. Therefore, according 
to the large $N_c$ QCD the VEV of the scalar field 
must obey the following expansion: 
\eq\label{VEV_Nc}
X_0(z) \to \frac{\sqrt{N_c}}{2} \, \hat{m} z 
+ \frac{1}{2\sqrt{N_c}} \, \Sigma z^3 \,, 
\en 
where $\Sigma \sim N_c$. 
It means that the contribution of the VEV 
of the scalar field $g X_0(z)$ scales as ${\cal O}(N_c)$. 
Inclusion of $\psi_+$ and $\psi_-$ mixing modifies 
the EOM for the fermion fields~(\ref{EOM_psi_pm}) as 
\eq\label{EOM_psi_mod}
\biggl[ i\not\!\partial + \gamma^5\partial_z
+ 2 A^\prime(z) \gamma^5
\mp \frac{e^{A(z)}}{R} \Big(m + \varphi(z)\Big) \biggr] 
\psi_{\pm,\tau}(x,z) = g X_0(z) \psi_{\mp,\tau}(x,z) \,. 
\en
After straightforward calculations we get the following EOMs for 
the $f^{L/R}_{\tau, n}(z)$ profiles: 
\eq\label{eq_KK_mod} 
\biggl[ -\partial_z^2
+ \frac{e^{2A}}{R^2} (m+\varphi)^2
\mp \frac{e^{A}}{R} \Big(A^\prime (m+\varphi) +
 \varphi^\prime\Big) \mp \frac{g X_0(z)}{\tilde M_{n\tau} \mp g X_0(z)} 
+ g^2 X_0^2(z)\biggr] \tilde f^{L/R}_{\tau, n}(z) 
= \tilde M_{n\tau}^2 \, \tilde f^{L/R}_{\tau, n}(z) \,.
\en 
Here symbol ``tilde'' on the top of the solutions 
$\tilde f^{L/R}_{\tau, n}(z)$ and $\tilde M_{n\tau}$ means 
that they differ from the set 
$f^{L/R}_{\tau, n}(z)$ and $M_{n\tau}$ in case when we neglect 
the mixing of $\Psi_+$ and $\Psi_-$ fermion fields. 
One can see that inclusion of the $g X_0(z)$ term gives a deviation of the mass 
spectra from a Regge-like trajectory. Lets estimate the contribution 
of $g X_0(z)$ term perturbatively using the solutions obtained for 
the case of a pure dilaton contribution (see Eqs.~\ref{bulk_prof_dilaton}). 
In particular, the nucleon mass shift $\Delta M_N$ 
due to the $g X_0(z)$ term is given by 
\eq 
\Delta M_N  = \sum\limits_\tau c_\tau \, \Delta M_{n\tau} \,, 
\en 
where 
\eq 
\Delta M_{n\tau} = \frac{g}{2} \, 
\int\limits_0^\infty \frac{dz}{z} \, X_0(z) \, 
\biggl[ \, 
(f^L_{\tau, n}(z))^2 - (f^R_{\tau, n}(z))^2 \, 
\biggr] = \frac{g}{\sqrt{N_c}} \, \frac{\Sigma}{4 \kappa^2} \,.  
\en 
One can see, that the term with current quark mass exactly vanishes 
and $\Delta M_{n\tau}$ does not depend on twist $\tau$. 
Therefore, using our condition $\sum\limits_\tau \, c_\tau = 1$ 
we get for the shift of the nucleon mass: 
\eq 
\Delta M_N = \frac{g}{\sqrt{N_c}} \, \frac{\Sigma}{4 \kappa^2} 
\sim N_c^0 \,. 
\en 
As we stressed before, the contribution of the VEV 
scalar field with dimension 3 is suppressed in comparison to 
the dilaton contribution encoding the VEV 
of the scalar field with dimension 2 by a factor $1/N_c$. 
 
Using a typical value for the quark condensate 
$\Sigma = (0.225 \ \mathrm{GeV})^3$ and the value of $\kappa = 0.383$ GeV 
we get the estimate of 
\eq 
\Delta M_N = \frac{g}{\sqrt{N_c}} \, 0.02 \, \mathrm{GeV} \,. 
\en
Then taking the typical values of the coupling $g \simeq 10$ 
using in hard-wall approaches~\cite{Baryons_ADS_QCD1} we finally get 
$\Delta M_N \simeq 115$~MeV for $N_c = 3$. 

\section{$P$--, $C$-- and $T$--parity transformations 
of bulk fields in AdS space} 

We use the chiral representation for the four-dimensional Dirac matrices 
$\gamma^\mu$ and $\gamma^5$: 
\eq 
\gamma^0 \;=\; 
\left(
\begin{array}{rl}
      0 &1\\
      1 &0 
\end{array}
\right) 
\,, \quad 
\gamma^i \;=\; 
\left(
\begin{array}{rl}
      0         &\sigma^i\\
     -\sigma^i  &       0
\end{array}
\right) 
\,, \quad 
\gamma^5 \;=\; 
\left(
\begin{array}{rl}
        -1 &0\\
         0 &1
\end{array}
\right).
\en 
The $(\frac{1}{2},0)$ left- and $(0,\frac{1}{2})$ 
right-handed Weyl spinors $\psi^{L/R}(x) 
= \frac{1 \mp \gamma^5}{2} \psi(x)$ are eigenstates of the chirality 
operator: 
\eq 
\gamma^5 \psi^{L/R} = \mp \psi^{L/R} 
\en 
and they form the $(\frac{1}{2},0) \oplus (0,\frac{1}{2})$  
Dirac bispinor $\psi = (\psi^L, \psi^R)^T$.  

The $P$-parity transformations of the fermion fields $\psi^{L/R}$, $\psi$ 
are defined as (here $U_P$ stands for the unitary operator of $P$-parity 
transformation): 
\eq 
& &U^{-1}_P \psi^{L/R}(t,\vec{x}) U_P 
= \gamma^0 \psi^{R/L}(t,-\vec{x})
\quad  \mathrm{and} \quad 
U^{-1}_P \psi(t,\vec{x}) U_P = \gamma^0 \psi(t,-\vec{x}) \,, \nonumber\\
& &U^{-1}_P \bar\psi^{L/R}(t,\vec{x}) U_P 
= \bar\psi^{R/L}(t,-\vec{x}) \gamma^0 
\quad  \mathrm{and} \quad 
U^{-1}_P \bar\psi(t,\vec{x}) U_P = \bar\psi(t,-\vec{x}) \gamma^0\,.  
\en   
Next we define the $P$-parity transformations of the five-dimensional bulk 
fields ($1/2$-fermion, vector and axial fields; in case of $1/2$-fermion 
fields we drop the summation over radial quantum number): 
\eq
U^{-1}_P \Psi_{\tau,\pm}(t,\vec{x},z) U_P &=& \pm \gamma^0 \gamma^5 \, 
\Psi_{\tau,\mp}(t,-\vec{x},z)\label{P_parity_fs1}\,, \nonumber\\ 
U^{-1}_P \bar\Psi_{\tau,\pm}(t,\vec{x},z) U_P &=& 
\pm \bar\Psi_{\tau,\mp}(t,-\vec{x},z) \, \gamma^0 \gamma^5 
\label{P_parity_fs2}\,, \nonumber\\ 
U^{-1}_P \Big(V^0(t,\vec{x},z),V^i(t,\vec{x},z),0\Big) U_P 
&=&  
\Big(V^0(t,-\vec{x},z),-V^i(t,-\vec{x},z),0\Big)\,, 
\nonumber\\ 
U^{-1}_P  \Big(A^0(t,\vec{x},z),A^i(t,\vec{x},z),0\Big) U_P 
&=&  
- \Big(A^0(t,-\vec{x},z),-A^i(t,-\vec{x},z),0\Big)\,. 
\en 
From the above equations we get the following conditions between 
the bulk profiles of fermion fields: 
\eq\label{Relation_FRL}
F^R_\pm(z) = \mp F^L_\mp(z) \,. 
\en  
Using the transformation of bulk fields it is easy to demonstrate that the 
effective Lagrangian/action of our model is $P$-parity invariant 
(some terms transform among themselves).  
In particular, we get 
\eq 
U^{-1}_P  \, \bar\Psi_{\pm,\tau}(t,\vec{x},z) 
\, \Psi_{\pm,\tau}(t,\vec{x},z) \, U_P
&=& - \bar\Psi_{\mp,\tau}(t,-\vec{x},z) 
\, \Psi_{\mp,\tau}(t,\vec{x},z)\,, \nonumber\\ 
U^{-1}_P  \, \bar\Psi_{\pm,\tau}(t,\vec{x},z) 
\, \hat{\cal D}_\pm(z) \, \Psi_{\pm,\tau}(t,\vec{x},z) \, U_P
&=& \bar\Psi_{\mp,\tau}(t,-\vec{x},z) 
\, \hat{\cal D}_\mp(z) \, \Psi_{\mp,\tau}(t,\vec{x},z)\,, \nonumber\\ 
U^{-1}_P  \, \bar\Psi_{\pm, \tau}(t,\vec{x},z) \, 
\hat{\cal V}_\pm(t,\vec{x},z) 
\, \Psi_{\pm, \tau}(t,\vec{x},z) \, U_P 
&=&  \bar\Psi_{\mp, \tau}(t,-\vec{x},z) \, \hat{\cal V}_\mp(t,-\vec{x},z) 
\, \Psi_{\pm, \tau}(t,-\vec{x},z) \,, \nonumber\\
U^{-1}_P  \, \bar\Psi_{\pm, \tau}(t,\vec{x},z) \, 
\hat{\cal A}_\pm(t,\vec{x},z) 
\, \Psi_{\pm, \tau}(t,\vec{x},z) \, U_P 
&=& \bar\Psi_{\mp, \tau}(t,-\vec{x},z) \, \hat{\cal A}_\mp(t,-\vec{x},z) 
\, \Psi_{\pm, \tau}(t,-\vec{x},z) \,, 
\en 
and therefore 
\eq 
U^{-1}_P \, {\cal L}_{\Psi}(t,\vec{x},z) \, U_P &=& 
{\cal L}_{\Psi}(t,-\vec{x},z)\,, \nonumber\\ 
U^{-1}_P \, {\cal L}_{V+A}(t,\vec{x},z) \, U_P &=& 
{\cal L}_{V+A}(t,-\vec{x},z)\,, \nonumber\\ 
U^{-1}_P \, {\cal L}_{\rm int}(t,\vec{x},z) \, U_P &=& 
{\cal L}_{\rm int}(t,-\vec{x},z)\,, \nonumber\\ 
U^{-1}_P \, S \, U_P &=& S \,,  
\en 
where $S$ is the effective action of our approach. 
Note, in the consideration of the vector (axial) field we apply 
the axial gauge $V(A)_z(x,z) = 0$.

Charge conjugation of four-dimensional spinors, vector 
and axial fields is defined 
with the use of the corresponding unitary operator $U_C$ as: 
\eq
U^{-1}_C \psi(x) U_C &=& C \, \bar\psi^T(x)\,, \hspace*{.9cm}\quad 
U^{-1}_C \bar\psi(x) U_C \ = \ \psi^T(x) \, C\,, \nonumber\\ 
U^{-1}_C \psi_{L/R}(x) U_C &=& C \, \bar\psi^T_{R/L}(x)\,, \quad 
U^{-1}_C \bar\psi_{L/R}(x) U_C \ = \ \psi^T_{R/L}(x) \, C\,, \nonumber\\ 
U^{-1}_C V_\mu(x) U_C &=& - V_\mu(x)\,, \hspace*{.85cm}\quad 
U^{-1}_C A_\mu(x) U_C \ = \ A_\mu(x)\,, 
\en 
where 
\eq 
C = i \gamma^0 \gamma^2\,, \quad C^T = C^\dagger = C^{-1} = - C
\,.  
\en 
$C$-transformations of AdS fields read as 
\eq 
U^{-1}_C \psi_\pm(x,z) U_C &=& \mp \, C \, \gamma^5 \, \bar\psi^T_\mp(x,z)\,, 
\hspace*{.25cm}\quad 
U^{-1}_C \bar\psi_\pm(x,z) U_C \ = \ \pm \psi^T_\mp(x,z) \gamma^5 C\,,
\nonumber\\ 
U^{-1}_C V_M(x,z) U_C &=& - V_M(x,z)\,, \quad \hspace*{.7cm}
U^{-1}_C A_M(x,z) U_C \ = \ A_M(x,z)\,.  
\en 
Therefore, one can straightforwardly prove that all terms of the effective 
action are $C$-invariant when the relation for the bulk profiles 
(\ref{Relation_FRL}) holds, e.g. 
\eq 
U^{-1}_C \bar\psi_{\pm}(x,z) \psi_{\pm}(x,z) U_C 
&=& - \bar\psi_{\mp}(x,z) \psi_{\mp}(x,z)\,, \nonumber\\ 
U^{-1}_C \bar\psi_{\pm}(x,z) \Gamma^M V_M(x,z) \psi_{\pm}(x,z) U_C 
&=& \bar\psi_{\mp}(x,z) \Gamma^M V_M(x,z) \psi_{\mp}(x,z)\,, \nonumber\\ 
U^{-1}_C \bar\psi_{\pm}(x,z) \Gamma^M A_M(x,z) \psi_{\pm}(x,z) U_C 
&=& - \bar\psi_{\mp}(x,z) \Gamma^M A_M(x,z) \psi_{\mp}(x,z)\,, \nonumber\\ 
U^{-1}_C \bar\psi_{\pm}(x,z) \Gamma^M i \Gamma^z V_M(x,z) 
\psi_{\pm}(x,z) U_C &=& - \,  \bar\psi_{\mp}(x,z) 
\Gamma^M i \Gamma^z V_M(x,z) \psi_{\mp}(x,z)\,, \nonumber\\ 
U^{-1}_C \bar\psi_{\pm}(x,z) \Gamma^M i \Gamma^z A_M(x,z) 
\psi_{\pm}(x,z) U_C 
&=& \bar\psi_{\mp}(x,z) \Gamma^M i \Gamma^z A_M(x,z) \psi_{\mp}(x,z)\,,  
\en 
etc. We therefore have $U^{-1}_C \, S \, U_C \ = \ S$. 

The $T$-parity transformation of four-dimensional spinors, vector and 
axial fields is defined 
with the use of corresponding antiunitary operator $U_T$ as: 
\eq
U^{-1}_T \psi(t,\vec{x}) U_T &=& T \, \psi(-t,\vec{x})\,, \hspace*{.9cm}\quad 
U^{-1}_T \bar\psi(t,\vec{x}) U_T \ = - \bar\psi(-t,\vec{x}) \, T\,, 
\nonumber\\ 
U^{-1}_T \psi_{L/R}(t,\vec{x}) U_T &=& T \, \psi_{L/R}(-t,\vec{x})\,, 
\quad 
U^{-1}_T \bar\psi_{L/R}(t,\vec{x}) U_T \ = \ 
- \bar\psi_{L/R}(-t,\vec{x}) T \,, \nonumber\\ 
U^{-1}_T \Big(V^0(t,\vec{x}),V^i(t,\vec{x})\Big) U_T 
&=&  
\Big(V^0(-t,\vec{x}),-V^i(-t,\vec{x})\Big)\,, 
\nonumber\\ 
U^{-1}_T \Big(A^0(t,\vec{x}),A^i(t,\vec{x})\Big) U_T 
&=&  
\Big(A^0(-t,\vec{x}),-A^i(-t,\vec{x})\Big)\,, 
\en 
where 
\eq 
T = - \gamma^1 \gamma^3\,, \quad T^T = T^\dagger = T^{-1} = - T\,.  
\en 
The $T$-transformations of the AdS fields read as                   
\eq
U^{-1}_T \psi_\pm(t,\vec{x},z) U_T &=& T \, \psi_\pm(-t,\vec{x},z)\,, 
\hspace*{.9cm}\quad 
U^{-1}_T \bar\psi_\pm(t,\vec{x},z) U_T \ = 
- \bar\psi_\pm(-t,\vec{x},z) \, T \,, 
\nonumber\\
U^{-1}_T \Big(V^0(t,\vec{x},z),V^i(t,\vec{x}),0\Big) U_T 
&=&  
\Big(V^0(-t,\vec{x},z),-V^i(-t,\vec{x},z),0\Big)\,, 
\nonumber\\ 
U^{-1}_T \Big(A^0(t,\vec{x},z),A^i(t,\vec{x},z),0\Big) U_T 
&=&  
\Big(A^0(-t,\vec{x},z),-A^i(-t,\vec{x},z),0\Big)\,. 
\en 
Therefore, one can straightforwardly prove that all terms of the effective 
action are separately $T$-invariant and $U^{-1}_T  \, S  \, U_T \ = \ S.$

\section{Structure integrals in the AdS/QCD model} 

The $C_i$ and $D_i$ functions defining the nucleon form factors 
are given by:   
\eq
C_1(Q^2) &=& D_1(Q^2) \ = \ \sum\limits_\tau \, c_\tau \, 
B(a+1,\tau) \, \left(\tau + \frac{a}{2}\right) \,, \nonumber\\ 
C_2(Q^2) &=& - D_2(Q^2) \ = \  
\frac{a}{2} \, \sum\limits_\tau \, c_\tau \, B(a+1,\tau) \,, \nonumber\\ 
C_3(Q^2) &=& a \, \sum\limits_\tau \, c_\tau \, B(a+1,\tau+1) \, 
\frac{a (\tau - 1) - 1}{\tau} 
\,, \nonumber\\ 
C_4(Q^2) &=& \frac{2 m_N}{\kappa} \,  
\sum\limits_\tau \, c_\tau \, (a+1+\tau) \, 
B(a+1,\tau+1) \, \sqrt{\tau - 1} \,, \nonumber\\ 
D_3(Q^2) 
&=& a \, \sum\limits_\tau \, c_\tau \, B(a+1,\tau+1) \, 
\frac{a (\tau - 1) + 2 \tau^2 - 1}{\tau} \,, 
\en 
where $a = Q^2/(4\kappa^2)$ and 
\vspace*{-.2cm} 
\eq 
B(m,n) = \frac{\Gamma(m) \, \Gamma(n)}{\Gamma(m+n)} 
\en  
is the Beta function. 

The slopes 
\eq 
C(D)_i^\prime(0) = \frac{dC(D)_i(Q^2)}{dQ^2}\bigg|_{Q^2=0} 
\en 
of the $C(D)_i$ functions are given by 
\eq
C_1^\prime(0) &=& D_1^\prime(0) \ = \  
- \frac{1}{8\kappa^2}  
\sum\limits_\tau \, c_\tau \, 
\biggl[ \frac{1}{\tau} + 2 \Big(\psi(\tau) - \psi(1)\Big) \biggr]\,, 
\nonumber\\ 
C_2^\prime(0) &=& - D_2^\prime(0) \ = \  
\frac{1}{8\kappa^2}  
\sum\limits_\tau \, \frac{c_\tau}{\tau}\,, 
\nonumber\\ 
C_3^\prime(0) &=& - \frac{1}{4\kappa^2}  
\sum\limits_\tau \, \frac{c_\tau}{\tau (\tau + 1)}\,, 
\nonumber\\ 
C_4^\prime(0) &=& - \frac{m_N}{2\kappa^3}  
\sum\limits_\tau \, c_\tau \, \sqrt{\tau-1} \, 
\biggl[ \frac{1}{\tau} +  \psi(\tau) - \psi(1) \biggr]\,, 
\nonumber\\ 
D_3^\prime(0) &=& \frac{1}{4\kappa^2}  
\sum\limits_\tau \, c_\tau \, \frac{2\tau^2 - 1}{\tau (\tau + 1)}\,, 
\en 
where 
\eq 
\psi(\tau) = 
\frac{d \log(\Gamma(\tau))}{d\tau} = 
\frac{1}{\Gamma(\tau)} \, \frac{d\Gamma(\tau)}{d\tau}
\en 
is the Digamma function obeying the recurrence formula 
$\psi(\tau + 1) = \psi(\tau) + 1/\tau \,.$

It is easy to see that the functions $C_i(Q^2)$ and $D_i(Q^2)$ 
scale at $Q^2 \to \infty$ as: 
\eq 
C_1^{\rm asym}(Q^2) & = & C_2^{\rm asym}(Q^2) \ = \ 
D_1^{\rm asym}(Q^2) \ = \ - D_2^{\rm asym}(Q^2) \ = \  
\frac{1}{2} \sum\limits_\tau \, c_\tau \, 
\frac{\Gamma(\tau)}{a^{\tau-1}}  \,, 
\nonumber\\ 
C_3^{\rm asym}(Q^2) &=& D_3^{\rm asym}(Q^2) \ = \ 
- \, \sum\limits_\tau \, c_\tau \, 
\frac{\Gamma(\tau) \, (\tau-1)}{a^{\tau-1}} \,, \nonumber\\ 
C_4^{\rm asym}(Q^2) &=& 
\frac{2m_N}{\kappa}   \, \sum\limits_\tau \, c_\tau \, 
\frac{\Gamma(\tau+1)}{a^\tau} \sqrt{\tau - 1} \,. 
\en 
From these considerations it is clear that the leading twist $\tau=3$
contributions to the nucleon form factors scale as 
\eq 
& &F_1^N(Q^2) \sim \frac{1}{Q^4}\,, \quad 
   F_2^N(Q^2) \sim \frac{1}{Q^6}\,, \quad 
   G_A(Q^2)   \sim \frac{1}{Q^4}\,. 
\en

\newpage

\begin{center}
\vspace*{1cm}
\epsfig{figure=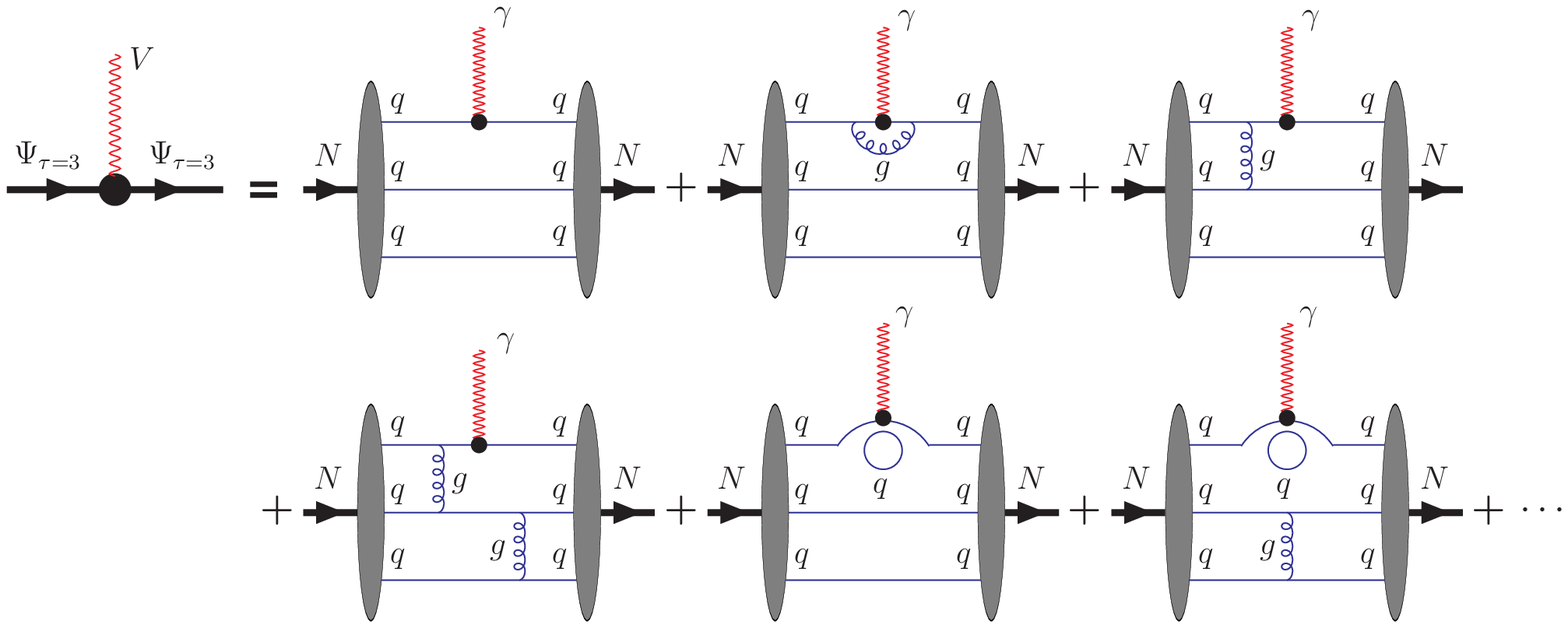,scale=.7}
\end{center}
\vspace*{0.5cm}
\noindent
{\bf Fig.1} Gauge/gravity duality between the vector-current transition 
matrix element involving twist dimension-3 fermion fields in AdS and the 
electromagnetic matrix elements involving twist-3 partonic Fock states 
in nucleons. 

\vspace*{.5cm}

\begin{center}
\vspace*{1cm}
\epsfig{figure=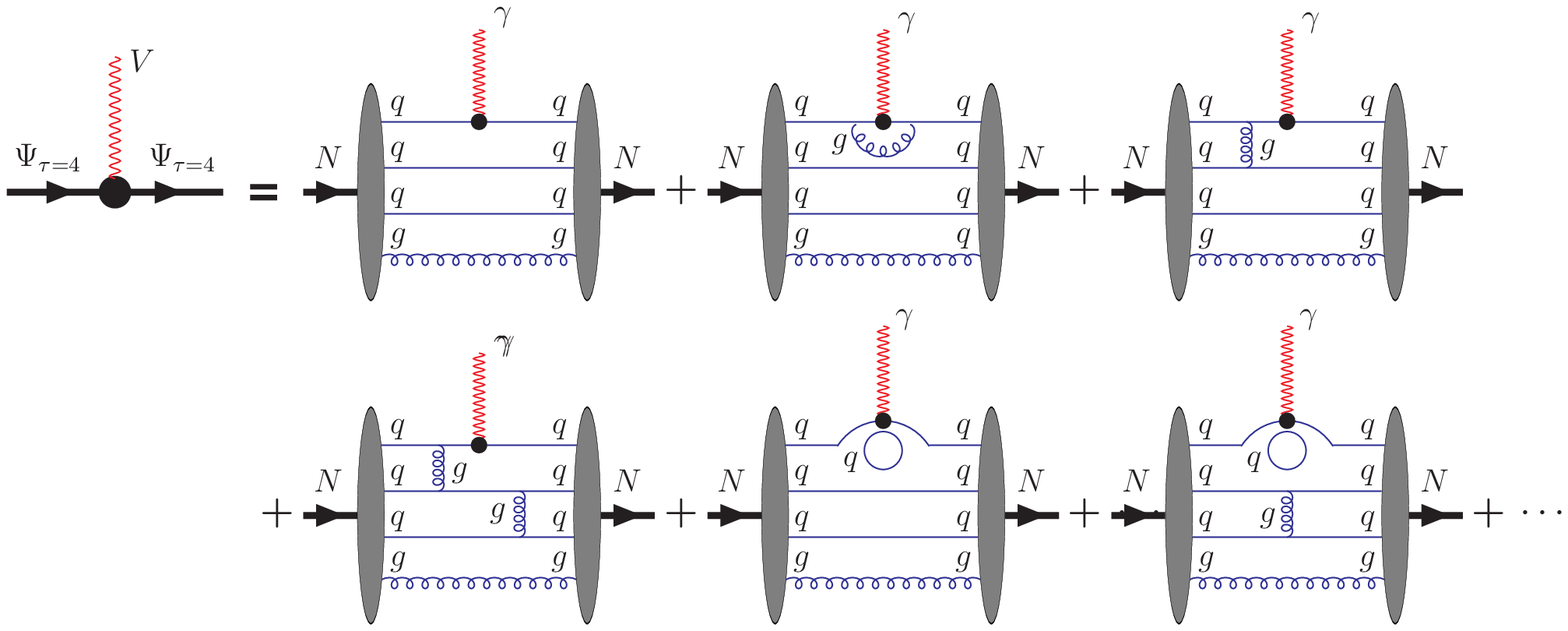,scale=.7}
\end{center}
\vspace*{0.5cm}
\noindent
{\bf Fig.2} Gauge/gravity duality between the vector-current transition 
matrix element involving twist dimension-4 fermion fields in AdS and the
electromagnetic matrix elements involving twist-4 partonic Fock states 
in nucleons. 

\vspace*{.5cm}

\begin{center}
\vspace*{1cm}
\epsfig{figure=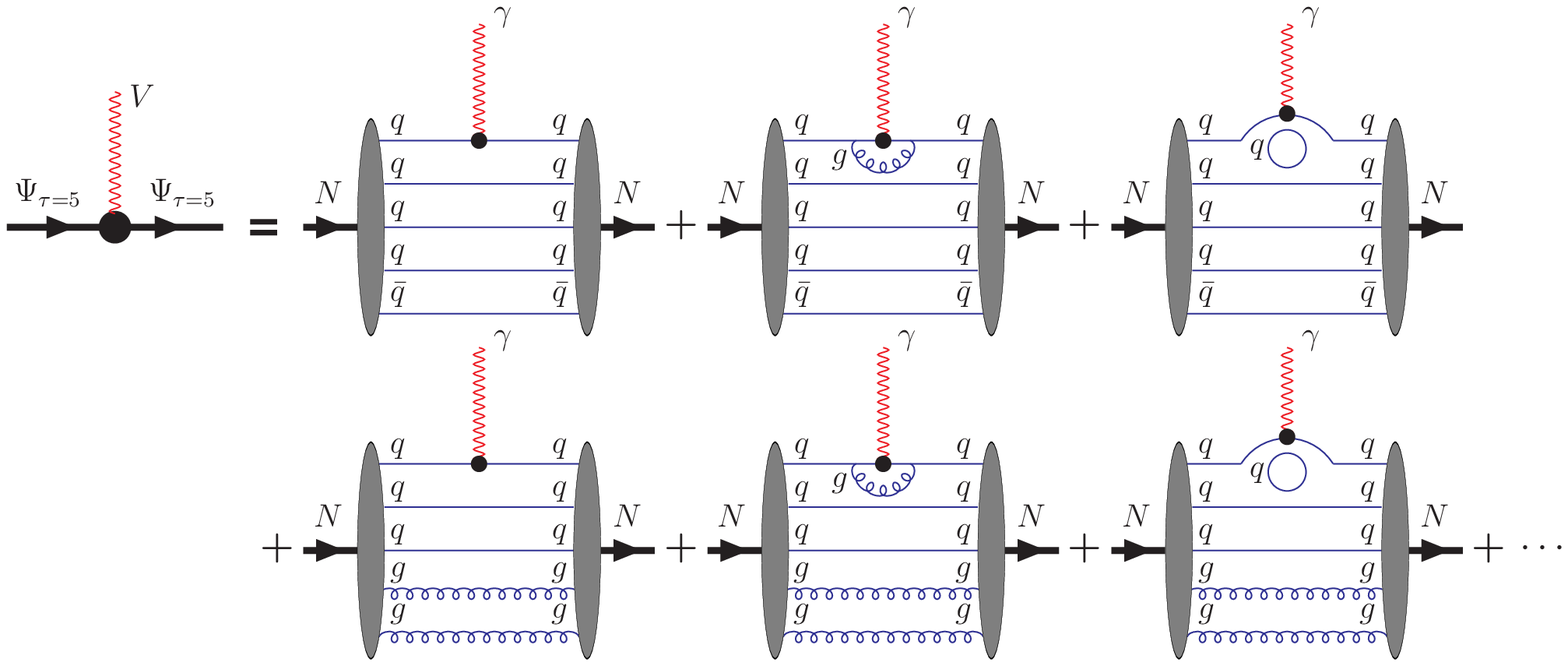,scale=.7}
\end{center}
\vspace*{0.5cm}
\noindent
{\bf Fig.3} Gauge/gravity duality between the vector-current transition 
matrix element involving twist dimension-5 fermion bulk fields in AdS and the 
electromagnetic matrix elements involving twist-5 partonic Fock states 
in nucleons. 

\newpage 

\begin{table}
\begin{center}
\caption{Mass and electromagnetic properties of nucleons} 

\vspace*{.25cm}

\def\arraystretch{1.5}
    \begin{tabular}{|c|c|c|}
      \hline
Quantity & Our results & Data~\cite{PDG} \\
\hline
$m_p$ (GeV)                &  0.93827     &  0.93827            \\
\hline
$\mu_p$ (in n.m.)          &  2.793       &  2.793              \\
\hline
$\mu_n$ (in n.m.)          & -1.913       & -1.913              \\
\hline
$g_A$                      & 1.270        & 1.2701              \\
\hline 
$r_E^p$ (fm)     &  0.840 &  0.8768 $\pm$ 0.0069 \\ 
\hline
$\la r^2_E \ra^n$ (fm$^2$) & -0.117 & -0.1161 $\pm$ 0.0022 \\ 
\hline
$r_M^p$ (fm)     &  0.785 &  0.777  $\pm$ 0.013 $\pm$ 0.010 \\ 
\hline
$r_M^n$ (fm)     &  0.792 &  0.862$^{+0.009}_{-0.008}$     \\
\hline
$r_A$ (fm)       &  0.667 &  0.67$\pm$0.01     \\
\hline
\end{tabular}
\end{center}
\end{table} 

\newpage 

\begin{figure} 
\begin{center}
\vspace*{1.25cm}
\epsfig{figure=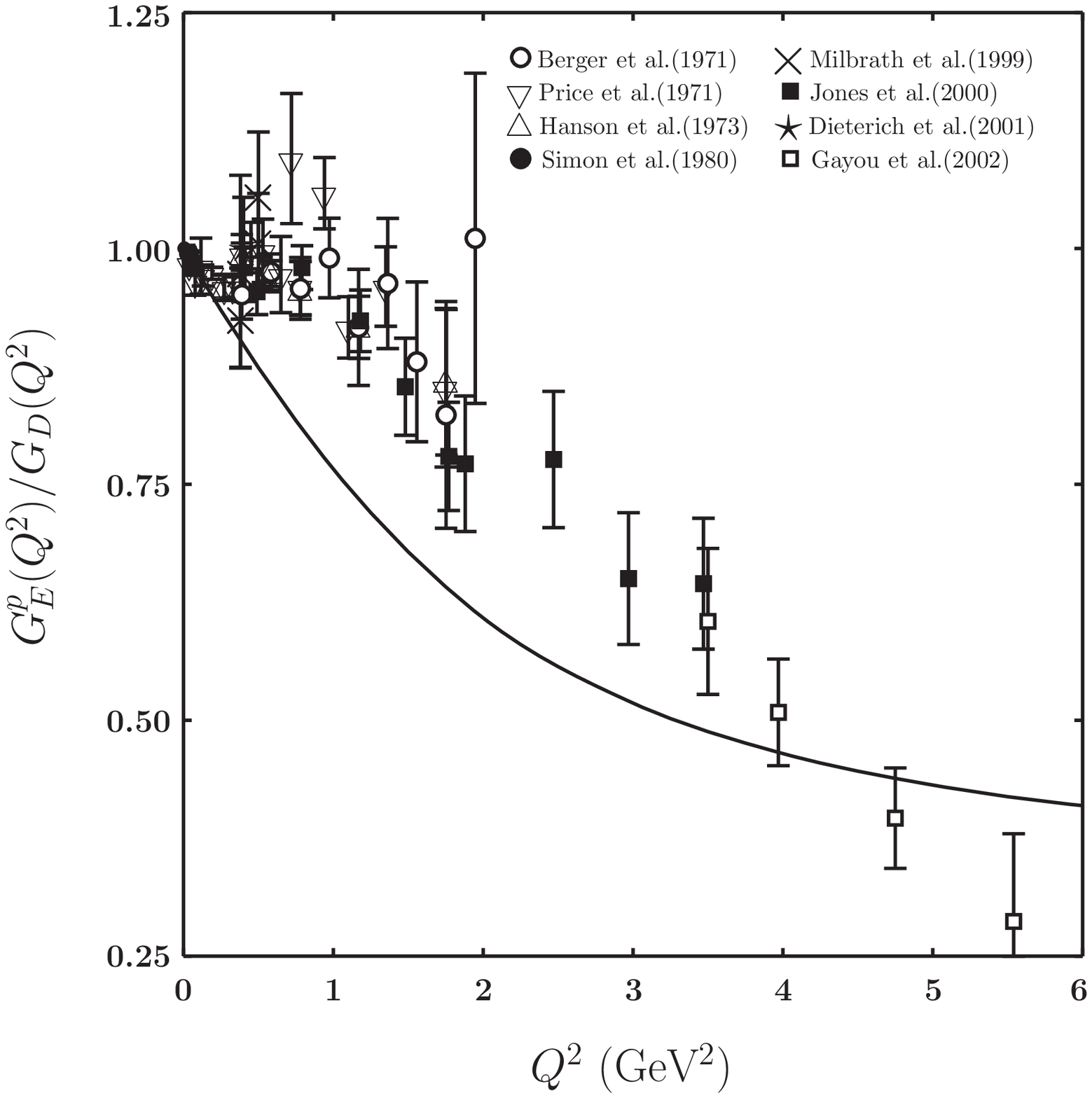,scale=0.6}
\end{center}
\vspace*{-2cm}
{\bf Fig.4:} Ratio $G_E^p(Q^2)/G_D(Q^2)$.  
Experimental data are taken from 
Refs.~\cite{Simon:1980hu,Price:1971zk,Berger:1971kr,Hanson:1973vf, 
Milbrath:1997de,Dieterich:2000mu,Jones:1999rz,Gayou:2001qd}. 

\vspace*{.5cm}

\begin{center}
\vspace*{1.25cm}
\epsfig{figure=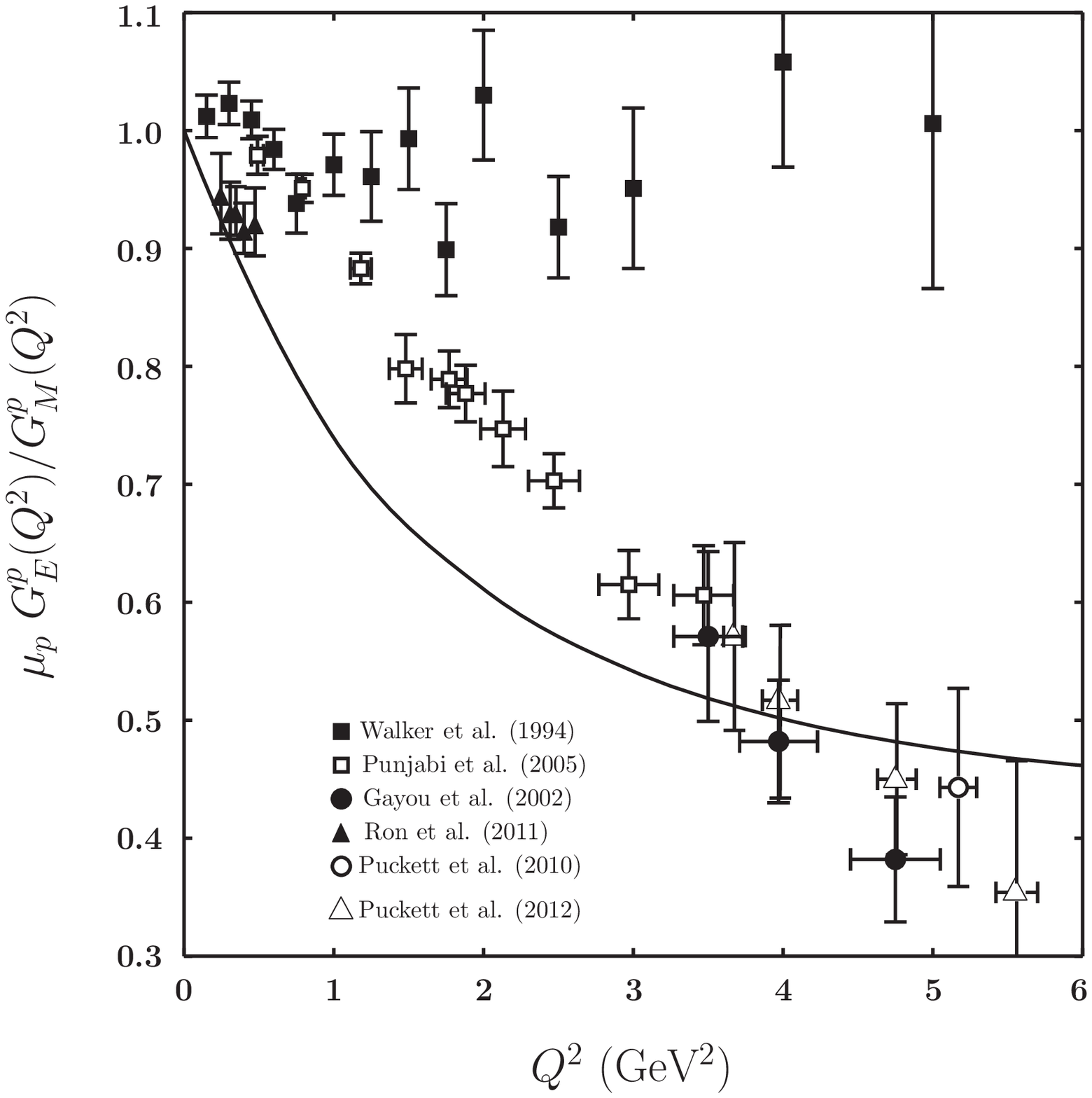,scale=0.6}
\end{center}
\vspace*{-2cm}
{\bf Fig.5:} Ratio $\mu_p G_E^p(Q^2)/G_M^p(Q^2)$ in comparison to the
experimental data taken from
Refs.~\cite{Walker:1993vj,Gayou:2001qd,Punjabi:2005wq,Ron:2011rd,%
Puckett:2011xg}. 
\end{figure} 

\newpage 

\begin{figure} 
\begin{center}
\vspace*{1.25cm} 
\epsfig{figure=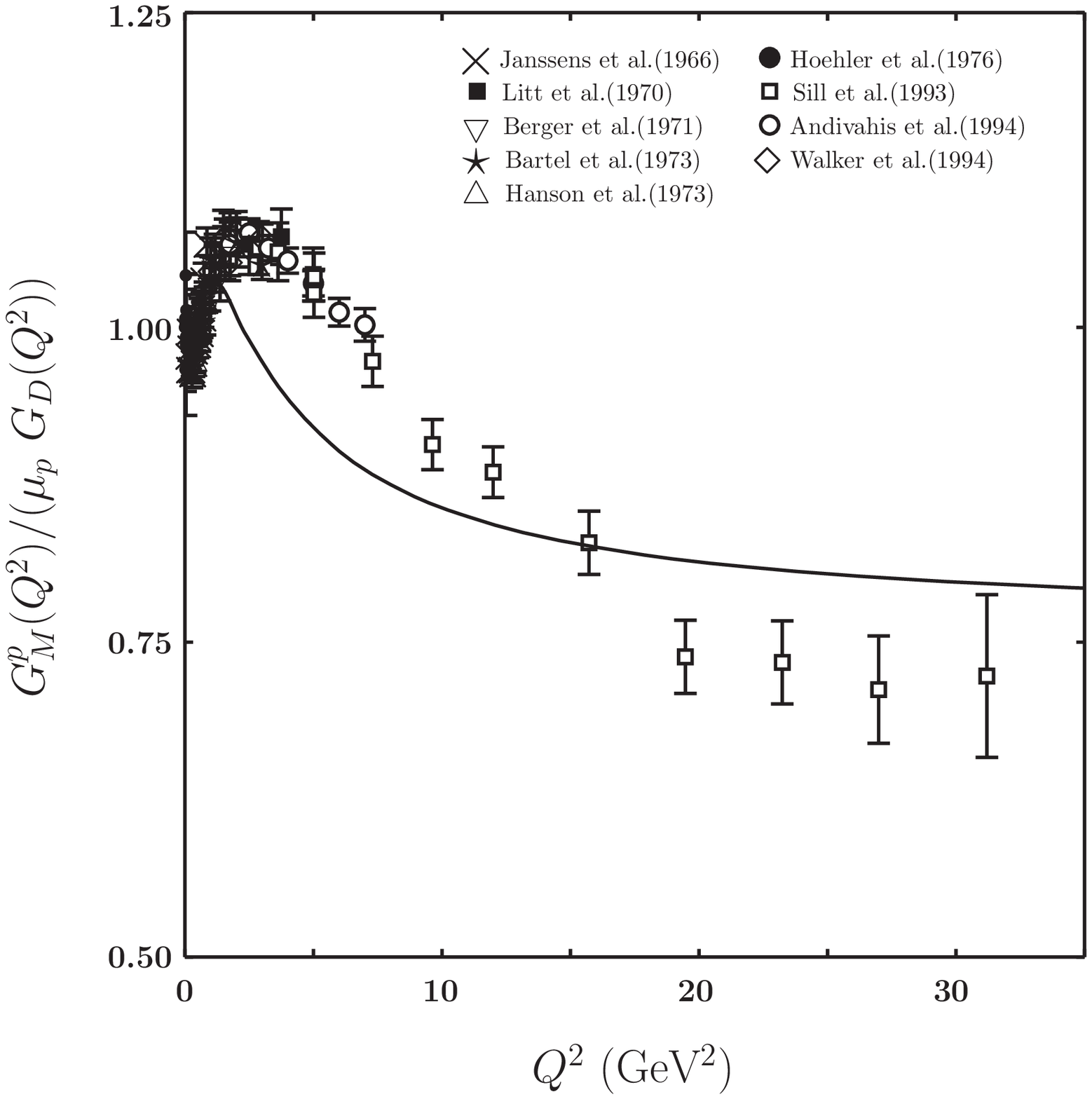,scale=0.6}
\end{center}
\vspace*{-2cm}
{\bf Fig.6:} Ratio $G_M^p(Q^2)/(\mu_p G_D(Q^2))$. 
Experimental data are taken from
Refs.~\cite{Walker:1993vj,Gayou:2001qd,Punjabi:2005wq}. 

\vspace*{.5cm}
\begin{center}
\vspace*{1.25cm} 
\epsfig{figure=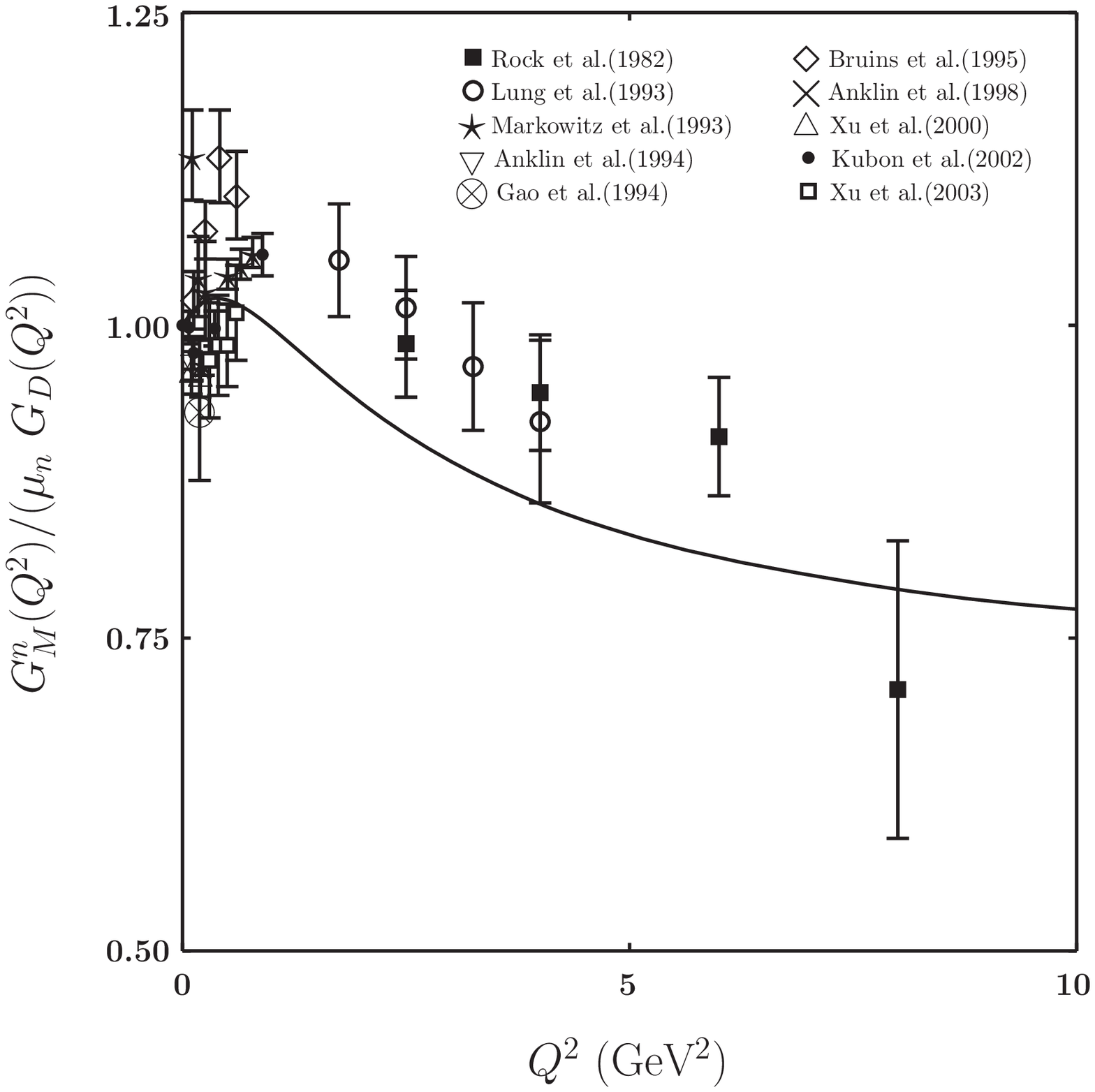,scale=0.6} 
\end{center}
\vspace*{-2cm}
{\bf Fig.7:} Ratio $G_M^n(Q^2)/(\mu_n\,G_D(Q^2))$. 
Experimental data are taken from 
Refs.~\cite{Lung:1992bu,Kubon:2001rj,Xu:2000xw,Anklin:1994ae,%
Anklin:1998ae,Rock:1982gf,Markowitz:1993hx,Bruins:1995ns,%
Xu:2002xc,Gao:1994ud}. 
\end{figure} 

\newpage 

\begin{figure} 
\begin{center}
\vspace*{1.25cm} 
\epsfig{figure=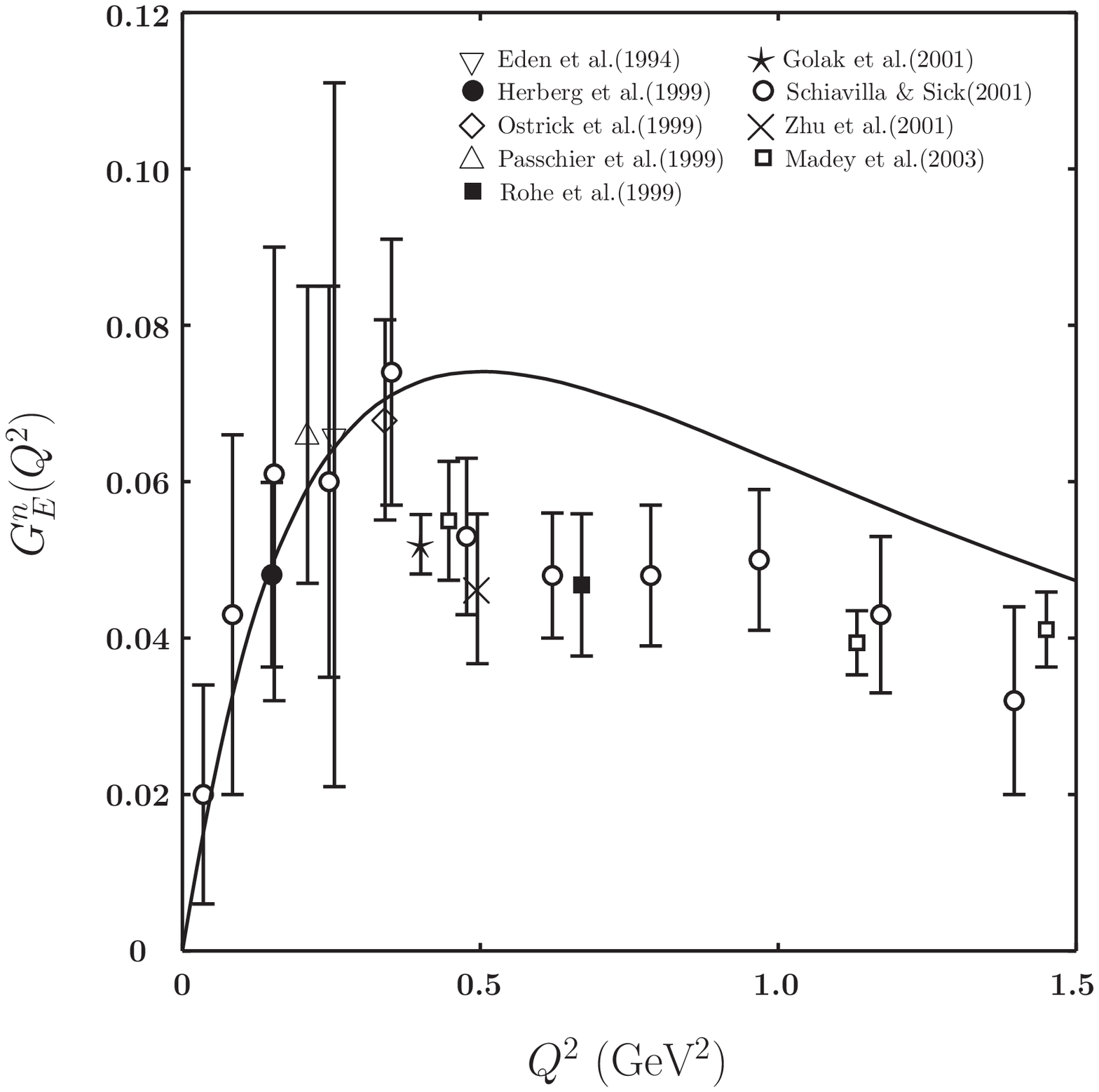,scale=0.6}
\end{center}
\vspace*{-2cm}
{\bf Fig.8:} 
The charge neutron form factor $G_E^n(Q^2)$. 
Experimental data are taken from Refs.~\cite{Herberg:1999ud,%
Passchier:1999cj,Eden:1994ji,Ostrick:1999xa,Golak:2000nt,%
Madey:2003av,Zhu:2001md,Rohe:1999sh,Schiavilla:2001qe}.

\vspace*{.5cm}
\begin{center}
\vspace*{1.25cm} 
\epsfig{figure=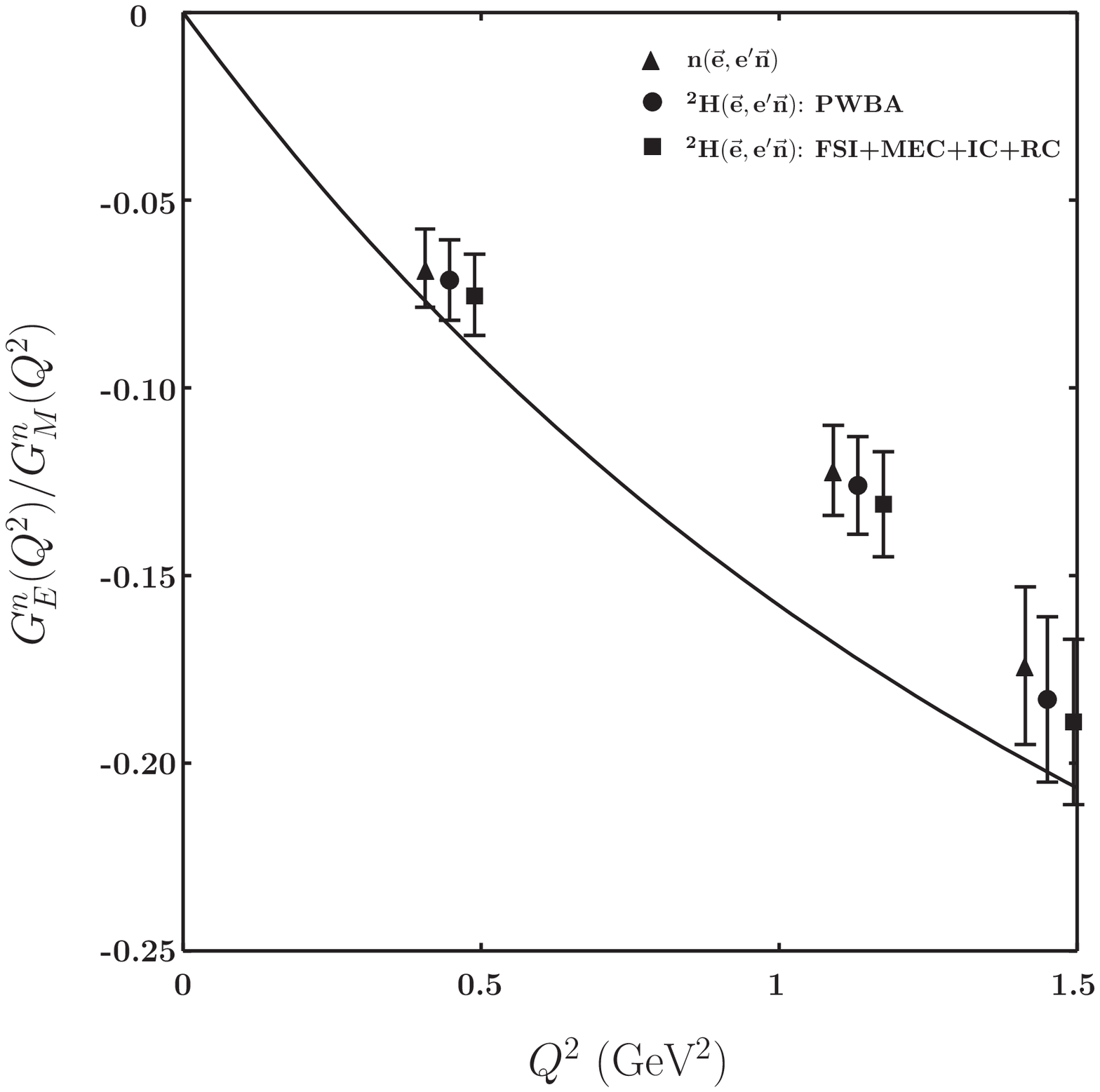,scale=0.6} 
\end{center}
\vspace*{-2cm}
{\bf Fig.9:} Ratio $G_E^n(Q^2)/G_M^n(Q^2))$. 
Experimental data are taken from 
Refs.~\cite{Plaster:2005cx}. 
\end{figure} 

\newpage 

\begin{figure} 
\begin{center}
\vspace*{1.25cm} 
\epsfig{figure=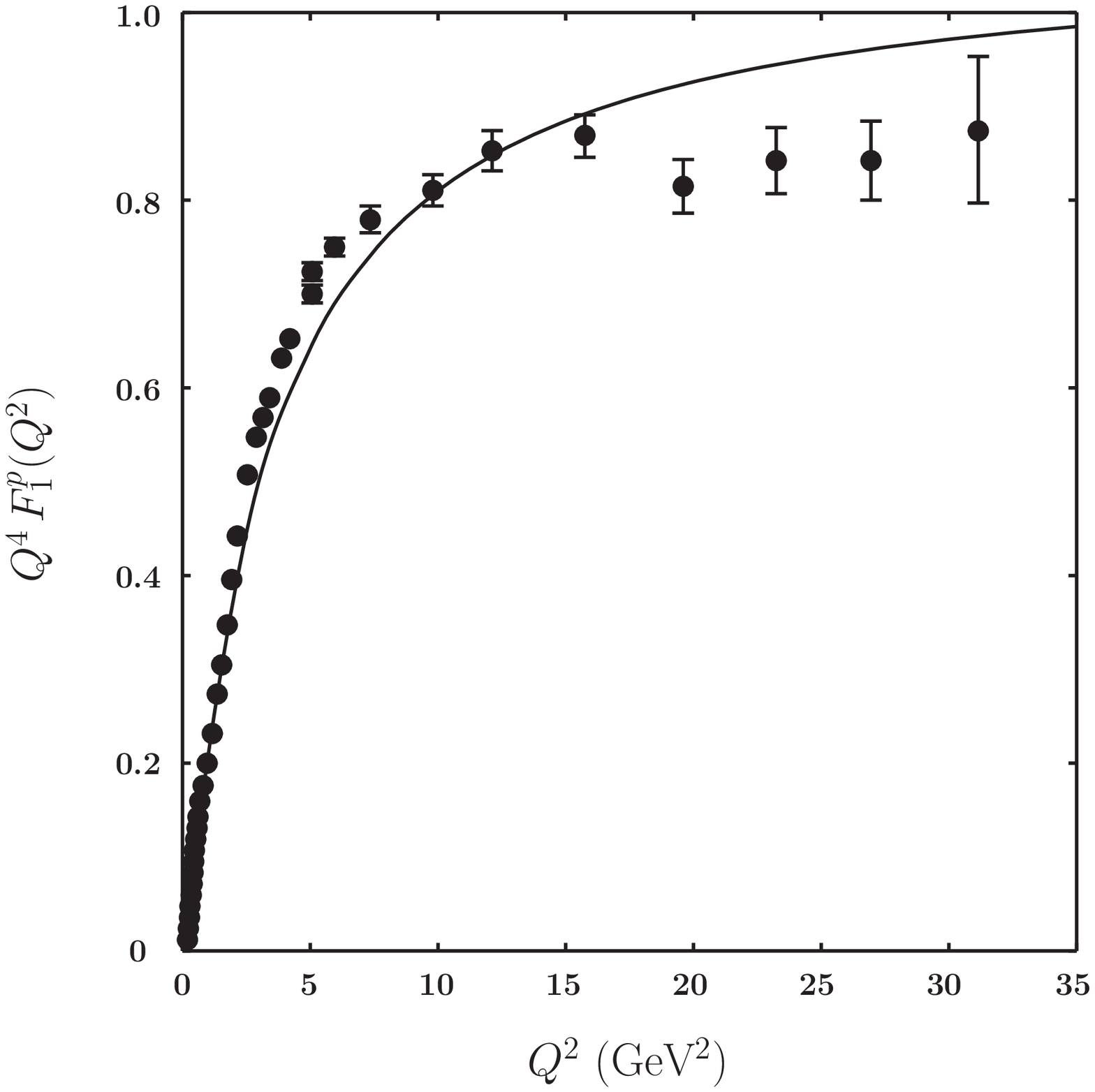,scale=0.6}
\end{center}
\vspace*{-2cm} 
{\bf Fig.10:} Proton Dirac form factor multiplied with $Q^4$.  
Experimental data are taken from Ref.~\cite{Diehl:2005wq}. 

\vspace*{.5cm}
\begin{center}
\vspace*{1.25cm} 
\epsfig{figure=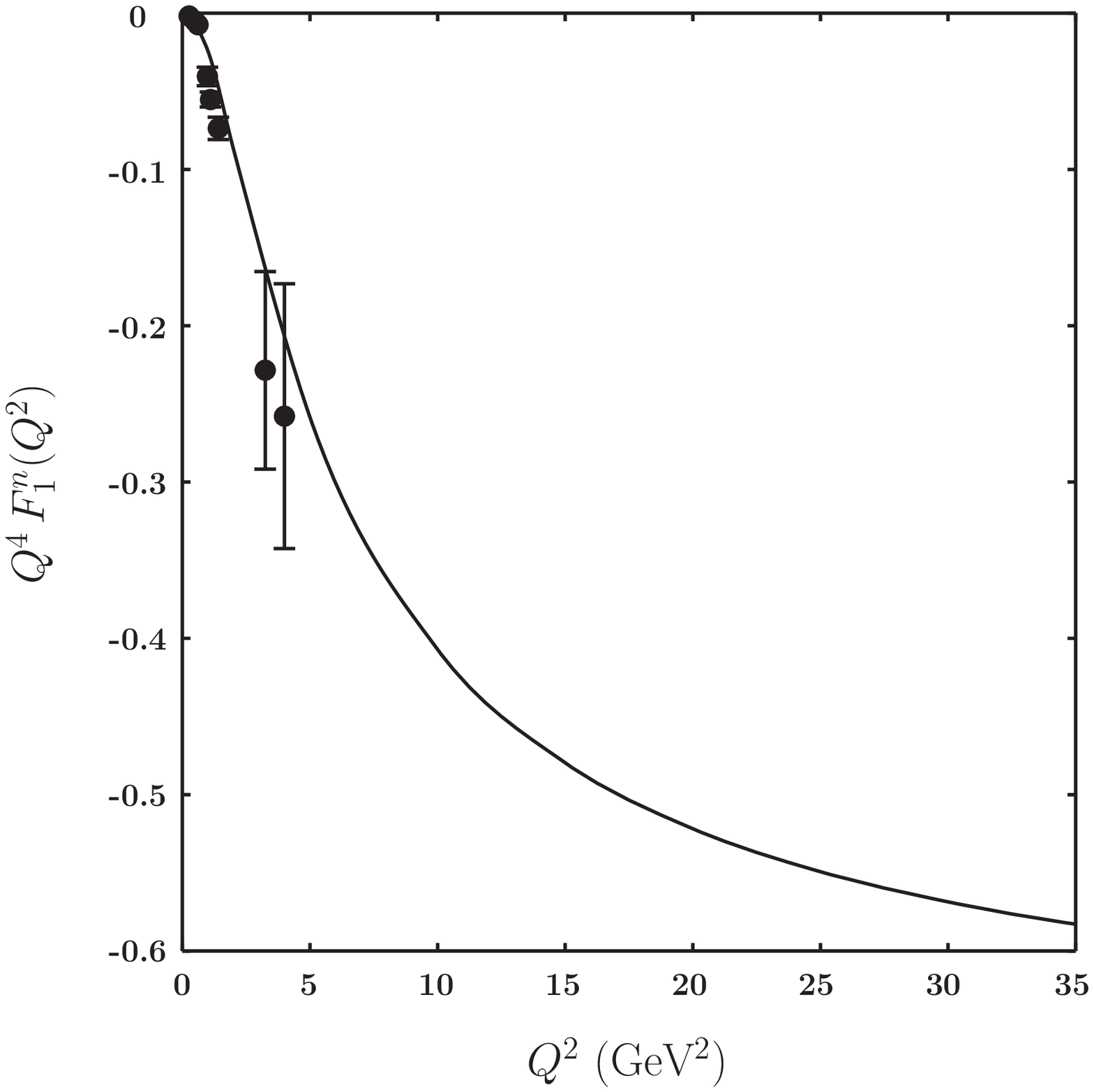,scale=0.6}
\end{center}
\vspace*{-2cm}
{\bf Fig.11:} Neutron Dirac form factor multiplied with $Q^4$.  
Experimental data are taken from Ref.~\cite{Diehl:2005wq}. 
\end{figure} 

\newpage 

\begin{figure} 
\begin{center}
\vspace*{1.25cm}
\epsfig{figure=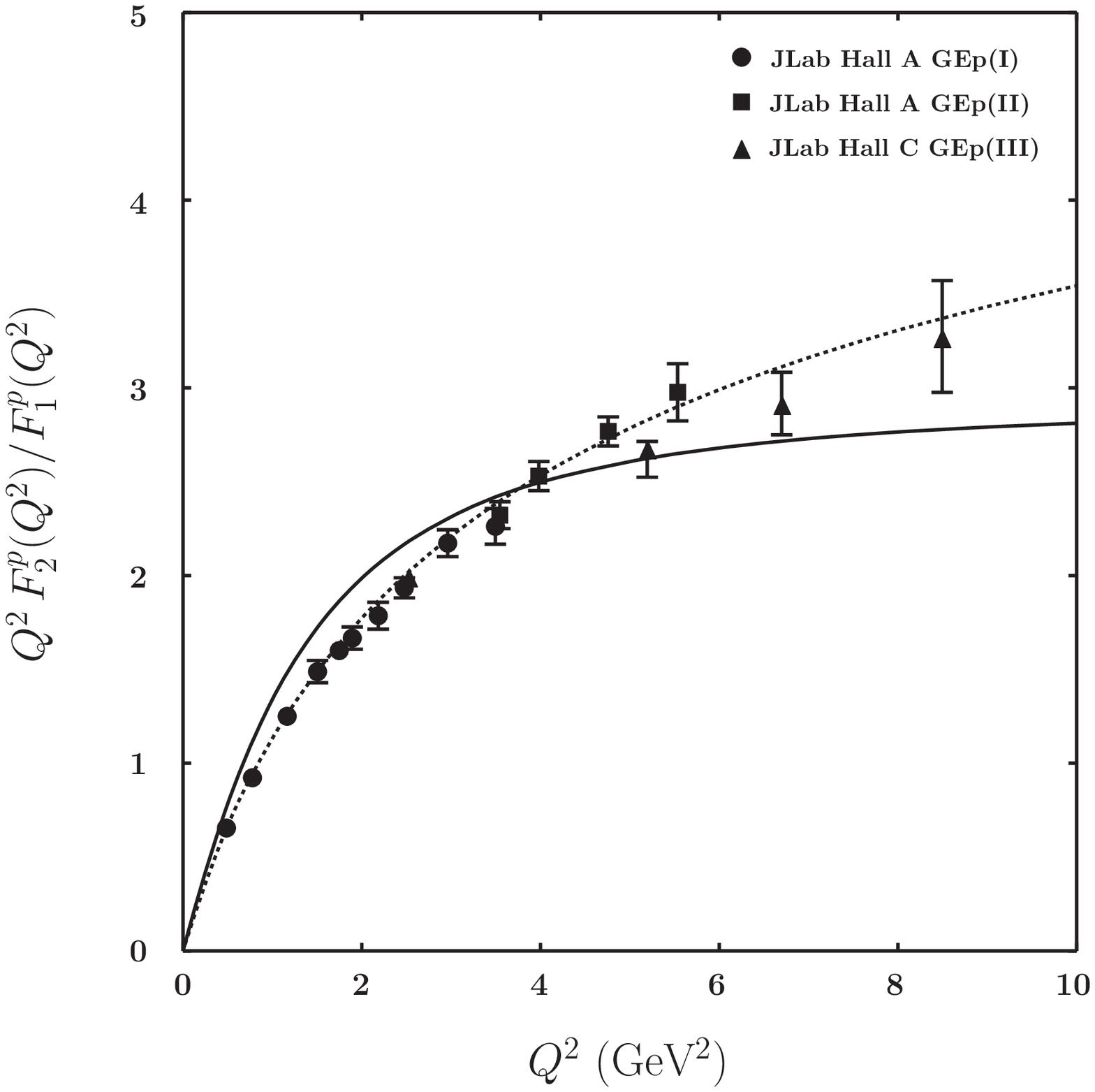,scale=0.6}
\end{center}
\vspace*{-2.2cm}
{\bf Fig. 12:} Results for $Q^2 F_2^p(Q^2)/F_1^p(Q^2)$. 
The solid line is the prediction of the soft-wall AdS/QCD model 
and the dashed line is the approximation of data suggested 
in Ref.~\cite{Brodsky:2002st}. Experimental data are taken
from Refs.~\cite{Jones:1999rz,%
Punjabi:2005wq,Puckett:2010ac,Crawford:2006rz,Jones:2006kf,%
Milbrath:1997de,MacLachlan:2006vw,Ron:2011rd,Zhan:2011ji}. 

\vspace*{.25cm}

\begin{center}
\vspace*{1.25cm}
\epsfig{figure=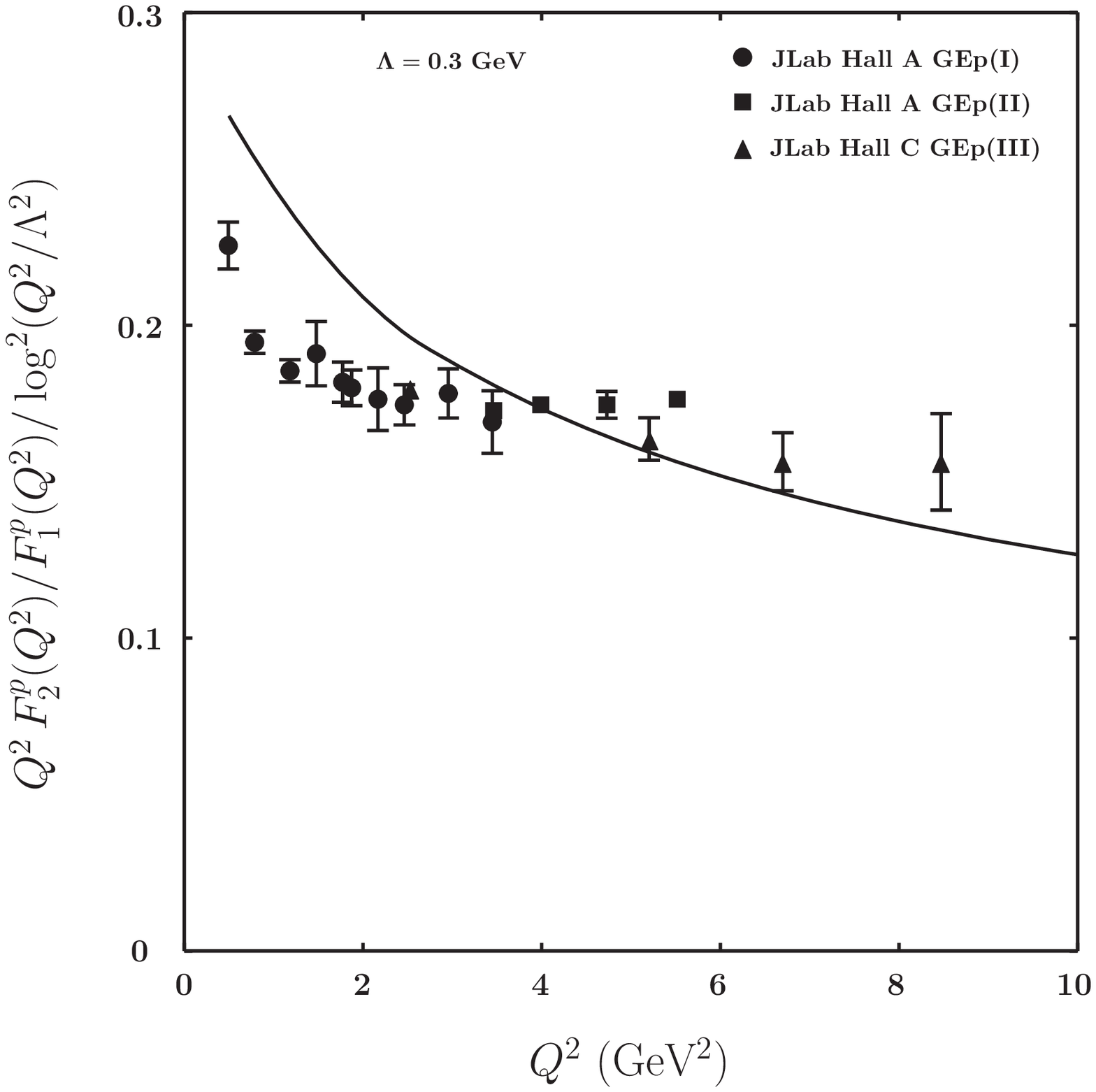,scale=0.6}
\end{center}
\vspace*{-2.2cm}
{\bf Fig. 13:} Results for $Q^2 F_2^p(Q^2)/F_1^p(Q^2)/\log^2(Q^2/\Lambda^2)$ 
at $\Lambda = 0.3$ GeV. 
Experimental data are taken from Refs.~\cite{Jones:1999rz,%
Punjabi:2005wq,Puckett:2010ac,Crawford:2006rz,Jones:2006kf,%
Milbrath:1997de,MacLachlan:2006vw,Ron:2011rd,Zhan:2011ji}. 
\end{figure} 

\newpage 

\begin{figure} 
\begin{center}
\vspace*{1.25cm}
\epsfig{figure=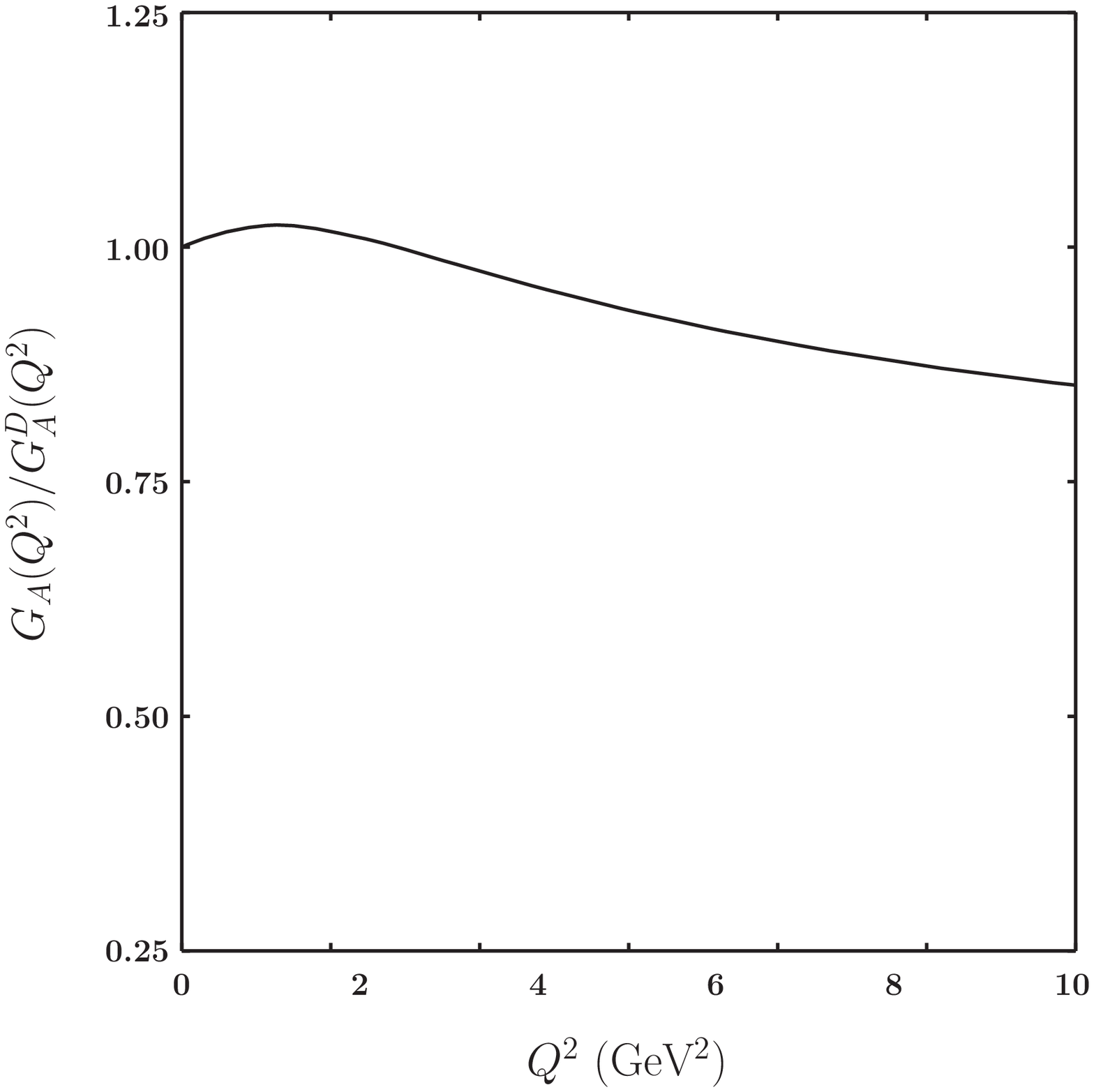,scale=0.6}
\end{center}
\vspace*{-2.2cm}
{\bf Fig.14:} Results for $G_A(Q^2)/G_A^D(Q^2)$.  
\end{figure} 

\end{document}